\documentclass[11pt, a4paper]{article}
\pdfoutput=1
\usepackage{jheppub}
\usepackage{mathdots, amsmath, amsfonts, amssymb}
\usepackage[latin1]{inputenc}

\usepackage{caption}
\usepackage{mathrsfs}
\usepackage{bbm}
\usepackage{epsfig}
\usepackage{latexsym}
\usepackage{slashed}
\usepackage{color}
\usepackage{amsthm}
\usepackage{amscd}
\usepackage{amstext}
\usepackage{enumerate}
\usepackage[section]{placeins}

\usepackage{hyperref}

\usepackage{graphics}

\newcommand{\beq}{\begin{equation}}
\newcommand{\eeq}{\end{equation}}
\newcommand{\bea}{\begin{eqnarray}}
\newcommand{\eea}{\end{eqnarray}}
\newcommand{\bi}{\begin{itemize}}
\newcommand{\ei}{\end{itemize}}
\newcommand{\ben}{\begin{enumerate}}
\newcommand{\een}{\end{enumerate}}

\newcommand{\dd}{\mathrm{d}}
\newcommand{\nn}{\nonumber}
\newcommand{\im}{\mathrm{i}}
\newcommand{\e}{\operatorname{e}}

\title{{\bf Holographic quenches and anomalous transport}}
\author[a]{Martin Ammon,}
\author[a]{Sebastian Grieninger,}
\author[a]{Amadeo Jimenez-Alba,}
\author[a]{Rodrigo P. Macedo}
\author[b]{and Luis
Melgar}

\affiliation[a]{{\it 
Theoretisch-Physikalisches Institut, Friedrich-Schiller University of Jena,}\\
{\it Max-Wien-Platz 1, 07743 Jena, Germany.}}
\affiliation[b]{{\it 
Blackett Laboratory, Imperial College,\\ 
Prince Consort Rd, London, SW7 2AZ, U.K..}
}

\emailAdd{martin.ammon@uni-jena.de}
\emailAdd{sebastian.grieninger@uni-jena.de}
\emailAdd{amadeo.jimenez.alba@uni-jena.de}
\emailAdd{rodrigo.panosso-macedo@uni-jena.de}
\emailAdd{luis.melgar@imperial.ac.uk}

\abstract{We study the response of the chiral magnetic effect due to continuous quenches induced by time dependent electric fields within holography. Concretely, we consider a holographic model with dual chiral anomaly and compute the electric current parallel to a constant, homogeneous magnetic field and a time dependent electric field in the probe approximation. We explicitly solve the PDEs by means of pseudospectral methods in spatial and time directions and study the transition to an universal ``fast'' quench response. Moreover, we compute the amplitudes, i.e.,~residues of the quasi normal modes, by solving the (ODE) Laplace transformed equations. We investigate the possibility of considering the asymptotic growth rate of the amplitudes as a well defined notion of initial time scale for linearized systems. Finally, we highlight the existence of Landau level resonances in the electrical conductivity parallel to a magnetic field at finite frequency and show explicitly that these only appear in presence of the anomaly. We show that the existence of these resonances induces, among others, a long-lived AC electric current once the electric field is switched off.}

\arxivnumber{1607.06817}

\begin{document}
\setcounter{tocdepth}{1}
\maketitle

\section{Introduction}
\noindent
In the last decade, there has been an increasing interest in macroscopic anomaly-driven effects. To date, several macroscopic phenomena have been shown to be related to the existence of quantum anomalies in the underlying microscopic theory. This includes, among others, the chiral magnetic (vortical) effect \cite{Fukushima:2008xe, Son:2009tf} which consists of the generation of an electromagnetic current in the direction of an external magnetic field (vortex), a massless propagating mode known as the chiral magnetic wave \cite{Kharzeev:2010gd}, and the enhancement of the electric conductivity in presence of parallel electromagnetic fields known as negative magnetoresistance~\cite{Nielsen:1983rb}. 

From the phenomenological point of view, the need for massless fermions strongly restricts the variety of experiments where these effects might be verified. To date the two main test grounds for anomalous transport are the state of matter known as quark gluon plasma (QGP) and Weyl/Dirac semimetals. In heavy ion collisions (HIC) the QGP is generated for $\sim 10^{-24}s$. In HIC very strong and transient magnetic fields are generated. Due to the high temperatures reached in the QGP, it is reasonable to neglect the quark masses. Moreover, experimental evidence \cite{Heinz:2008tv} shows that the QGP is, despite the high temperatures, strongly coupled. 

The second kind of systems in consideration is of a very different nature. Weyl (Dirac) semimetals are solid state crystals. These present effective Weyl (Dirac) fermionic degrees of freedom in concrete points of the Brioullin zone known as Weyl nodes. The main difference between Dirac and Weyl semimetals is that in the former the nodes for opposite chiralities coincide in momentum space while in the latter these are separated. Remarkably, one of the aforementioned effects, the ``negative magnetoresistance'', has already been measured in several Weyl and Dirac semimetals \cite{Li:2014bha, Li2015, 2015PhRvX...5c1023H, 2016NatCo...710301L}. Although it is usually said that Weyl semimetals are the 3D graphene, it is not yet clear whether the former are strongly coupled too. Therefore it is of special interest to study these systems from a strong coupling perspective. In this sense the gauge/gravity duality \cite{Maldacena:1997re} (for textbooks see \cite{Ammon:2015wua,Nastase,Schalm}) is a specially adequate theoretical tool to study these processes. 

 A complete study of anomalous transport effects in both HIC and condensed matter systems requires to address the question of how the system behaves out of equilibrium. Several studies in this direction include computations for fixed axial charge and dynamic magnetic fields at weak coupling to leading order in $\alpha_s$ \cite{Kharzeev:2009pj, Jimenez-Alba:2015bia}   and at strong coupling via holography \cite{Yee:2009vw, Landsteiner:2013aba}. The fate of the chiral magnetic effect (CME) in presence of time-dependent external electric and magnetic fields has been studied at weak coupling for concrete configurations of the electromagnetic fields \cite{Fukushima:2015tza, Iwazaki:2009bg}. Moreover, the authors of   \cite{Lin:2013sga} studied the evolution of the CME during thermalization in holography. It is desirable to have a complete description of anomalous transport for realistic systems.

 Most of the past effort in this direction had its focus on the phenomenology of the QGP in HIC. However, the appearance of Weyl/Dirac semimetals has affected the order of relevance of the different features to be addressed. An example of this is the axial charge time-dependence. In the field of QGP and HIC it has been argued \cite{Lin:2013sga, Kharzeev:2001ev, Hegde:2008rm} that the estimated decay rate of axial charge is of the order of the lifetime of the plasma and therefore as a first approximation the charge can be taken to be constant. In ``chiral'' condensed matter system, however, the story is quite different.
 
 Indeed, it is known that in these systems, the chiral magnetic current vanishes in equilibrium even at very large temperatures \cite{Basar:2013iaa}. This is due to the fact that the Weyl nodes are filled up to the same energy (the Fermi energy) and that the Nielsen-Ninomiya theorem \cite{nino} applies. For materials with the Weyl nodes placed at the same energy, one can generate a net chiral magnetic current by inducing a finite axial chemical potential. This is achieved by the non-conservation of the axial charge density in the presence of parallel electric and magnetic fields. Therefore, the mechanism studied here can be used to generate the CME in Weyl semimetals.\footnote{One can also consider shifting the relative energy of the Weyl nodes \cite{Cortijo:2016wnf}}
 
Motivated by these ideas, we focus in this paper on the time evolution of the CME in presence of time dependent axial charge. In order to introduce axial charge in the system in a time dependent fashion, we consider the anomaly expression in presence of external electromagnetic fields and in absence of axial currents  
  \begin{equation}\label{eq:anomH}
  \dot{\rho}_5 \sim \mathbf{E\cdot B}\,.
  \end{equation}
  This equation tells us that the evolution of the axial charge is determined by the time dependence of $\mathbf{E\cdot B}$. Concretely, we will consider a static magnetic field and a dynamic electric field.
  
  The possible strongly coupled nature of Dirac/Weyl semimetals motivates us to apply holography. This is not, however, the only reason to consider holography for this topic. As extensively discussed in the literature, holography is nicely suited to perform real time computations which, by means of the holographic dictionary, reduce to solving the bulk system of PDEs. As shown later, our case can be reduced to one hyperbolic PDE describing a quenched current operator in presence of a magnetic field. This connects the initial motivation of this work to the topic of (quantum) quenches in holography. 

In recent years, there has been a lot of interest in studying (quantum) quenches due to new experimental results. From the theoretical point of view, we have to 
compute the (quantum) evolution of an isolated system in the presence of a time-dependent parameter in the Hamiltonian. As explained above, for strongly coupled quantum field theories, gauge/gravity dualities are an ideal playground to tackle such questions. One remarkable outcome of previous studies of quenches in holography is the appearance of universal scaling behaviours in the ``early time'' response \cite{Buchel:2013lla} for fast quenches. We explore these ideas within our system and determine how the anomaly affects them.  

In order to study all these features, it is in principle necessary to solve the initial-boundary value problem i.e. solving PDEs. Indeed we explicitly solve the equations by means of pseudospectral methods. However, for linear systems there is a way out of this. Based on several works from the numerical GR community in flat space \cite{PhysRevD.45.2617, Nollert:1998ys, Ansorg:2016ztf}, we are able to numerically compute the residue of the QNMs of the system by just solving the ODEs that result from Laplace transforming the PDEs. In other words, one does not need to solve the explicit time dependent problem to obtain the residues. As shown in \cite{Ansorg:2016ztf}, generically, this information allows to determine the full response only from an initial time scale $\tau$ on. $\tau$ is obtained from the amplitudes and therefore depends not only on the system under consideration but on the initial and boundary conditions as well. When this applies, it means that the knowledge of \textit{all} QNM and their residues is not enough to determine the response of the system at initial times. We explore the applicability of these ideas in AdS and give a physical interpretation to $\tau$ from the dual theory point of view.  

The paper is structured as follows. In section \ref{sec:model}, we present and discuss our holographic model and the ansatz that we use for the different fields. This section includes some discussion regarding the 1-point function renormalization, gauge fixing and the definition of the current. In section \ref{sec:rodrigo}, we adapt the spectral decomposition analysis developed for asymptotically flat Schwarzschild in \cite{Ansorg:2016ztf} to AdS-Schwarzschild. Concretely, we explicit compute the amplitudes of the QNMs in our system for different boundary data. In section \ref{sec:initial}, we focus on the initial time behaviour of the current. First we look at the dependence of $\tau$ on the different parameters of the system. After this, following \cite{Buchel:2013lla}, we look at the response of the current for fast quenches and its dependence on the anomaly coefficient and the magnetic field. Finally,  in section \ref{sec:late} we study the late time response where we identify QNMs approaching the real axis with increasing magnetic field. This modes are responsible for a long lived oscillatory late time behaviour of the current and for a resonant behaviour under oscillatory sources. By computing the QNMs with backreaction, we argue that this is indeed a consequence of the axial anomaly in our class of holographic models. Moreover, we relate the QNMs to resonances due to Landau levels. We finish this work in section \ref{sec:discussion} with a discussion of the results and some hints on possible future directions. 
\section{The model}\label{sec:model}
\noindent
In this work we consider a $U(1)\times U(1)$ model, which has been previously applied to study anomalous transport in holography \cite{Gynther:2010ed, Jimenez-Alba:2014iia, Jimenez-Alba:2014pea}. None of these studies, however, has considered the effect that dynamical generation of axial charge may have in the response. Our aim here is to compute time dependent setups for the CME in holography. Concretely we focus on the evolution of the effect in presence of electric field driven axial charge generation. This is in contrast to the usual ``by hand'' implementation of net charge via (axial) chemical potential \cite{ Yee:2009vw, Gynther:2010ed}.  

The 5-dimensional matter Lagrangian consists of two photon fields $(A_\mu, V_\mu)$ coupled by a Chern-Simons (CS) term in the bulk 
\begin{equation}
\mathcal{L}= -\frac{1}{4}F^{\mu\nu}F_{\mu\nu}-\frac{1}{4}H^{\mu\nu}H_{\mu\nu}+ \frac{\kappa}{2}\varepsilon^{\mu\alpha\beta\rho\lambda}A_\mu\left(F_{\alpha\beta}F_{\rho\lambda}+
3H_{\alpha\beta}H_{\rho\lambda}\right)\,,
\end{equation}%
with $F=\dd A$, $H=\dd V$. Here, bulk indices are denoted by Greek letters running from $0$ to $4$. Computing the divergence of the dual current operators one can show that, on-shell
\begin{equation}\label{eq:ward}
\langle\partial_i J^i_V \rangle=0\,,\hspace{2cm}\langle\partial_i J^i_A \rangle=\frac{\kappa}{2}\tilde\varepsilon^{\,ijkl}\left( F_{ij}F_{kl}+3H_{ij}H_{kl} \right),
\end{equation}%
with $\tilde\varepsilon^{\,ijkl}$ being the epsilon symbol in the boundary theory and Latin indices running from $0$ to $3$. These Ward identities correspond to an abelian anomaly. The Bardeen counter term has been chosen such that only one of the currents is not conserved. It is therefore customary to identify $J_V$ and $J_A$ with the dual vector and axial currents, respectively. This allows us to meaningfully consider external ``electromagnetic'' fields and study the evolution of the system in such backgrounds. Moreover, the relative factor between both CS terms is chosen so that the relative factor in the r.h.s.~of equation \eqref{eq:ward} coincides with that of the abelian part of the anomaly in QCD \cite{Bertlmann:1996xk}.

As a first step towards solving the full non-linear system we consider the probe limit of the system. This limits the validity of the results. We comment on this later, when the results are discussed. As background metric we consider Schwarzschild-AdS$_5$. In infalling Eddington-Finkelstein coordinates the metric reads
\begin{equation}
\dd s^2=\frac{1}{\rho^2}\,\left(-f(\rho)\dd v^2-2\,\dd v \dd\rho +\dd x^2+\dd y^2+\dd z^2\right),\hspace{2cm}f(\rho)=1-\rho^4\,.
\end{equation}%
Note that the conformal boundary is located at $\rho=0$ and without loss of generality we have rescaled the coordinates such that the AdS radius is $L=1$ and the black-hole horizon is  at $\rho=1$.

The equations of motion for the gauge fields read
\begin{align}
 \nabla_\mu F^{\mu \nu}+\frac{3\kappa}{2}\varepsilon^{\nu\alpha\beta\rho\lambda}
 \left(F_{\alpha\beta}F_{\rho\lambda}+H_{\alpha\beta}H_{\rho\lambda}\right)=0\,,
  \nabla_\mu H^{\mu \nu}+3\kappa\,\varepsilon^{\nu\alpha\beta\rho\lambda}
 F_{\alpha\beta}H_{\rho\lambda}=0\,.
\end{align}

\subsection{Setup}\noindent
As explained in the introduction, in order to control the dynamic generation of axial charge we need a time dependent $\mathbf{E\cdot B}$. We keep the magnetic field static and just set a time dependent electric field. Therefore, we consider the following ansatz
\begin{align}\label{eq:ansatz}
 A_v(v, \rho),  \hspace{1cm}   V_y(x)=Bx,    \hspace{1cm}  V_z(v, \rho), 
 \end{align}
with boundary condition
\begin{align}\label{eq:conditions}
 \dot V_z(v, \rho\rightarrow 0)=E(v)\,,
 \end{align}
and the usual regularity conditions for $A_v(v,\rho)$ at the horizon. Here, the time derivate $\partial_v$ is represented by the dot. This corresponds to a time independent magnetic field and time dependent electric field both homogeneous and pointing in the $z$-direction in the boundary theory. Due to the anomaly, a non-trivial time component of the axial gauge field is needed to get equations for $V_z$ and $V_y$  consistent with the choices in equation \eqref{eq:ansatz}. The equations for this ansatz read

\begin{align}
 A_v''-\frac{1}{\rho}A_v'- 12\kappa B V_z' \rho=0\,,\label{eq:eom1}\\
  V_z''+ \left(\frac{f'}{f}-\frac{1}{\rho} \right)V_z' -\frac{2}{f}\dot V_z '+\frac{1}{\rho f}\dot V_z-12 \kappa
 B \frac{\rho}{f} A_v'=0\,,\label{eq:eom2}\\\
 \dot A_v'-12 \kappa B \rho  \dot V_z=0\,,\label{eq:eom3}
 \end{align}
with the notation $'$ for the radial derivative $\partial_\rho$. Integrating equation~(\ref{eq:eom3}) in time one obtains 
\begin{equation}\label{eq:av}
A_v'=12 \kappa B \rho   V_z+C_1(\rho)\,.
\end{equation}
Substituting this result back into equation \eqref{eq:eom1}, one finds $C_1(\rho)=\rho C$. We will later see in equation \eqref{eq:regularsolultions} that $C$ is just a gauge shift for $V_z$ and will fix it to $C=0$. We can then reduce the system of PDEs to a single hyperbolic PDE for the $z-$component of the vector field
\begin{align}\label{eq:vzeq}
  V_z''+ \left(\frac{f'}{f}-\frac{1}{\rho} \right)V_z' -\frac{2}{f}\dot V_z '+\frac{1}{\rho f}\dot V_z-\left(12 \kappa
 B \rho\right)^2\frac{1}{f} V_z-12 \kappa  B \rho^2\frac{C}{f} =0\,.
 \end{align}
The asymptotic boundary expansions read
\begin{align}\label{eq:expansion}
V_z&\sim V_{0}+\rho\, \dot V_{0} + \rho^2\,\tilde V +\frac{1}{2} \rho^2\,\log(\rho) \ddot V_{0}+ \mathcal{O}(\rho^3),\\
A_v&\sim A_{0}+\rho^2 \tilde A + \mathcal{O}(\rho^3).
\end{align}
In order to compute the dual correlators we need to specify boundary terms. We have  
\begin{equation}
-\int_{\partial \mathcal{M}} \dd^4x\, \sqrt{-\gamma}\, \frac{1}{4} F_{ij}F^{ij}  \log(\rho)-\int_{\partial \mathcal{M}}\dd^4x\,\sqrt{-\gamma}\, \frac{1}{4} H_{ij}H^{ij} \left( \log(\rho)-\frac{1}{2} \right),
 \end{equation}%
with $\gamma$ the induced metric on the boundary. The first term is an infinite counter-term  fixed by the theory. In addition, for convenience in our numerical calculations, we fix the renormalization scheme by introducing the second (finite) gauge invariant counter term. We make this choice to avoid explicit contributions of the logarithmic coefficient in \eqref{eq:expansion} to the one point function which in our scheme is given by 
\begin{equation}\label{eq:1pcons}
\langle J_z^V\rangle_\text{cons}=2 \tilde V - 12\kappa B A_{0}\,,
\end{equation}%
where the subscript stands for ``consistent'' current. From previous studies of the CME in holography it is known that this version of the current can be problematic \cite{Yee:2009vw, Gynther:2010ed} due to the difficulty to distinguish between the chemical potential and the dual coupling associated to $A_v$ in certain gauge. It is now well understood  \cite{Gynther:2010ed} that imposing a vanishing $A_v$ at the horizon gives rise to a vanishing \textit{consistent} current in the time independent limit. One way to circumvent these issues is to compute the \textit{covariant} current instead   
  \begin{equation}
  \langle J_z^V\rangle_\text{cov}=\langle J_z^V\rangle_\text{cons} + 12\kappa B A_{0}\,.
  \end{equation}%
 In Eddington-Finkelstein coordinates, however, regularity does not imply a vanishing gauge field at the horizon. Therefore we are free to choose a gauge in which the field vanishes at the boundary $A_v(\rho=0)=A_0=0$. With this choice the covariant and consistent currents are indistinguishable
\begin{equation}
\label{eq:Current}
\langle J_z \rangle\equiv\langle J_z^V \rangle= 2 \tilde V\,. 
\end{equation}
From now on we stick to this choice and omit the ``vector'' superscript. Then, our problem reduces to computing the coefficient of the normalisable mode of the $V_z$ field. To this end, we must solve the dynamical equation~\eqref{eq:vzeq} with the boundary condition \eqref{eq:conditions} and additional initial data.

In order to construct the initial data, we assume that our system is initially in equilibrium, i.e. described by a stationary configuration. Mathematically, this assumption implies that all time derivatives of the boundary data $V_0(v)$ must vanish at $v=0$. In practice, this condition is hardly realised. Nonetheless, as we are going to mention afterwards, we choose specific quenches profiles for the function $V_0(v)$ in such a way that 
$\dfrac{\dd^j}{\dd v^j} V_0(0)  \sim 0$.  

The time-independent equation \eqref{eq:vzeq} can be rewritten in terms of a new variable $u\equiv \rho^2$ as an inhomogeneous Legendre equation
\begin{align}\label{eq:leg}
 (1-u^2) V_z''- 2 u V_z' -36 \left( \kappa B \right)^2 V_z= 3 \kappa
  B \,C\,,
 \end{align}
with regular solutions
\begin{equation}\label{eq:regularsolultions}
V_z=C_2\mathcal{P}_l(\rho^2)-\frac{C}{12 \kappa B}\hspace{2cm} l=\frac{1}{2}\left(-1+\sqrt{1-144\,(\kappa B)^2} \right)\,.
\end{equation}%
Here, $\mathcal{P}_l$ are the l-th Legendre functions of the first kind. The constant $C$ is just a shift in the field. It becomes clear that $A_v'$ in equation~\eqref{eq:av} is $C$ independent. We will from now on gauge fix it to $C=0$. The constant $C_2$ determines the initial value of the current and it can be related to the initial value of the axial chemical potential 
\begin{equation}
\mu_A\equiv A_v(\rho=0)-A_v(\rho=1)=12\kappa B C_2 \int^0_1 \dd\rho \,\mathcal{P}_l(\rho^2)\,\rho\,.
\end{equation}%
Since the adiabatic regime implies that the equilibrium formula is valid at any time, the adiabatic response is given by the time independent solution. The asymptotic expansion of the stationary solutions in equation \eqref{eq:regularsolultions} read 
\begin{equation}\label{eq:legenderexpansion}
V_z\sim V_0- 18(\kappa B)^2\frac{\Gamma\left(\frac{1}{2}(l
+2) \right) \Gamma\left(\frac{1}{2}(1-l) \right)}{\Gamma\left(\frac{1}{2}(l
+3) \right)\Gamma\left(\frac{1}{2}(2-l) \right)}V_0 \rho^2+\mathcal{O}(\rho^3)\,,
\end{equation}%
with $\Gamma$ the Euler gamma functions. Therefore, in the adiabatic limit
\begin{equation}\label{eq:adiabatic}
\langle J_z \rangle= -36(\kappa B)^2\frac{\Gamma\left(\frac{1}{2}(l
+2) \right) \Gamma\left(\frac{1}{2}(1-l) \right)}{\Gamma\left(\frac{1}{2}(l
+3) \right)\Gamma\left(\frac{1}{2}(2-l) \right)} V_0(v)\, .
\end{equation}%
 Moreover, for $\kappa B\gg1$ it asymptotes as 
\begin{equation}\label{eq:adiabaticJ}
\langle J_z \rangle\sim  -12 \kappa B V_0(v)\,.
 \end{equation}%
 It is possible to understand the above equation in terms of the effect of the Chiral Magnetic Wave. If a system displays a massless mode with a finite residue $R_0$, the response to an externally applied electric field $E(v) =\dot V_0(v)$ has a contribution of the form $\langle J(\omega)  \rangle= R_0 E(\omega)/\omega$ in Fourier space. Hence the D.C. conductivity is infinite if $R_0(\omega=0) \ne 0$. For a given $V_0(v)$, it is then straightforward to see that $\langle J(v)  \rangle=  R_0 V_0(v)$. In our situation, $R_0 = -12 \kappa B$ for large $\kappa B$ and thus at low frequencies the D.C. conductivity scales as $\sigma(\omega) \sim - 12 \kappa B\,\delta (\omega)$.

In order to compute the current outside the adiabatic limit, we make use of the asymptotic expansion \eqref{eq:expansion} and introduce an auxiliary field $U(v,\rho)$ via
\begin{equation}
\label{eq:AuxField_U}
V_z(v,\rho) \equiv V_{0}(v)+\rho\, \dot V_{0}(v) + \rho^2\,U(v,\rho) +\frac{1}{2}\rho^2\,\log(\rho) \left(\ddot V_{0}(v)+\rho \dddot V_{0}(v) \right).
\end{equation}%
Substituting equation~\eqref{eq:AuxField_U} into the original equation \eqref{eq:vzeq} we obtain
\beq
\label{eq:DynEq_TimeDom}
\left[ - \rho\,(1-\rho^4)\, \frac{\partial^2 }{\partial \rho^2 } - \left( 3 - 7\rho^4\right)\, \frac{\partial}{\partial \rho} + (8 + \lambda^2)\,\rho^3 \right]U(v,\rho)  + \left[  2\,\rho\, \frac{\partial }{\partial \rho } +  3\ \right] \dot{U}(v,\rho) + S(v,\rho) = 0.
\eeq
with $\lambda = 12\,\kappa B$. By doing this, we automatically incorporate the information about the boundary condition \eqref{eq:conditions} into the inhomogeneity $S(v,\rho)$ given by
\beq
S(v,\rho) = a_4(\rho) \frac{\dd^4}{\dd v^4} V_0(v) + a_3(\rho) \frac{\dd^3}{\dd v^3} V_0(v) + a_2(\rho) \frac{\dd^2}{\dd v^2} V_0(v) + a_1(\rho) \frac{\dd}{\dd v} V_0(v) + a_0(\rho) V_0(v),
\eeq
and 
\bea
\label{eq:Source_ak}
a_4(\rho) &=& \frac{1}{2}\left(2\rho +  5\rho\log(\rho) \right), \quad a_3(\rho) = \frac{15}{2}\rho^4\log(\rho)+ \frac{\lambda^2}{2}\rho^4\log(\rho)+4\rho^4-1 \notag \\
a_2(\rho) &=& \rho^3\left(4\log(\rho)+3+\frac{\lambda^2}{2}\log(\rho) \right), \quad
a_1(\rho) = \left(3+\lambda^2\right)\rho^2, \quad  a_0(\rho) = \lambda^2\,\rho\,. 
\eea
Therefore, equation~\eqref{eq:DynEq_TimeDom} is to be solved as an initial value problem with
\beq
\label{eq:ID_U}
U(0,\rho) \equiv U_{\rm in}(\rho) = \frac{V_0}{\rho^2}\left[ \frac{P_l(\rho^2)}{P_l(0)} - 1\right].
\eeq
The most natural way to solve~\eqref{eq:DynEq_TimeDom} is to numerically integrate it in time. In this work, the time evolution is performed with the fully spectral code\footnote{In order to double check our results, we also evolve the equations with a Crank-Nicolson time integrator while keeping the spectral method in the spatial direction.} introduced in \cite{Macedo:2014bfa} (the numerical details are described in appendix \ref{sec:methods}). We use this highly accurate numerical method to obtain the full response of the current operator. In particular, it provides us with a reference solution, against which we can compare the results from the strategy based on the spectral decomposition of the solution $U(v,\rho)$ in terms of the quasi-normal modes. 

\section{QNM amplitudes from Laplace analysis in AdS}\label{sec:rodrigo}
As stated in the introduction, solving the PDE is not the only way to determine the response of the current. Given the spectral decomposition for the dual operator \begin{equation}\label{eq:spectraldec}
\langle J_z(v) \rangle=\sum_{n=0}^\infty A_n \e^{-\im\omega_n v}\, ,
\end{equation}%
solving the problem reduces to obtaining the eigenfrequencies $\omega_n$ and the amplitudes $A_n$ for given initial and boundary data. It is well known how to obtain the quasi-normal frequencies in holography \cite{Kovtun:2005ev}. However this is not the case for the amplitudes $A_n$. Of course a possibility is to fit the data obtained from the PDE to the previous formula (see for example \cite{Heller:2013oxa}). It is, nevertheless, possible to compute $A_n$ directly without need of explicitly solving the PDEs. Although this topic has a long history in the community of numerical general relativity \cite{PhysRevD.45.2617, Nollert:1998ys} only recently \cite{Ansorg:2016ztf} has it been shown how to compute these amplitudes in the case of asymptotically flat Schwarzschild black hole for rather generic initial data.

More specifically, the authors formulate the wave-equation describing the propagation of fields on the Schwarzschild background in terms of a particular coordinate system, where the surfaces of constant time are spacelike hypersurfaces that penetrate the black-hole horizon and extend to future null infinity. By using the standard framework provided by the Laplace-transformation, they develop a semi-analytical algorithm to obtain ``eigenvalues" and ``eigenvectors" (related only to the wave-equation in question) and amplitudes (related to the particular initial data being used). Most importantly, they introduce a well defined time scale $\tau$ for which the spectral decomposition in the form \eqref{eq:spectraldec} is valid, based on the growth rate of such amplitudes.  

In this section, we provide the first steps towards the application of these techniques in asymptotically AdS and we focus on the role that the $\tau$ may have in the dual theory.

\subsection{Laplace transformation}
We consider the compact form of the dynamical equation \eqref{eq:DynEq_TimeDom} for the function $U(v,\rho)$ 
\beq
 {\boldsymbol \alpha} [U] +  {\boldsymbol \beta} \left[ \frac{\partial} {\partial v} U\right] + S = 0,
\eeq
with given initial data $U_{\rm in}(\rho)$ specified in \eqref{eq:ID_U}. Here, ${\boldsymbol \alpha(\rho)}$ and ${\boldsymbol \beta(\rho)}$ are differential operators acting on the radial coordinate $\rho$, which can be easily read from \eqref{eq:DynEq_TimeDom}. The former is  second order, whereas the latter is first order. As mentioned, the source term $S(v,\rho)$ contains information about the boundary data $V_0(v)$ and is generically written in the form
\beq
S(v,\rho) = \sum_{i=0}^{N_S} a_i(\rho) \frac{\dd^i}{\dd v^i} V_0(v).
\eeq
Here, $N_S=4$ and the functions $a_i(\rho)$ are summarised in equations~\eqref{eq:Source_ak}. Applying the Laplace transformation\footnote{Note that the parameter $s$ of the Laplace transformation is related to the Fourier parameter by the relation $s=-\im \omega$.}
\beq
\bar{U}(\rho;s) = {\cal L}[U(v,\rho)](s) = \int\limits_0^{\infty} \\\dd v \, U(v,\rho) \e^{-sv}
\eeq
and taking into account that for a generic function $f(v,\rho)$
\beq
{\cal L}\left[\frac{\partial^{n}}{\partial v^n}f(v,\rho)\right](s) = s^n\,\bar{f}(\rho;s) - \sum_{k=0}^{n-1} s^{n-k-1} \frac{\partial^{k}}{\partial v^k}f(0,\rho),
\eeq
we obtain
\bea
 {\boldsymbol \alpha} [ \bar{U}] &+& s\, {\boldsymbol \beta} [ \bar{U}]  = {\boldsymbol \beta} [ U_{\rm in}] - \bar{S} \quad {\rm with} \label{eq:LapTransEq}\\
   \bar{S}(\rho;s) &=&  \bar{V}_0(s) \sum_{i=0}^{N_S} a_i(\rho)  s^i\, - \underbrace{ \sum_{i=0}^{N_S} \sum_{k=0}^{i-1} s^{i-k-1} \frac{\dd^{k}}{\dd v^k}V_0(0)}_{\sim \, 0}. \label{eq:LapTransSource}
\eea
As observed in the previous section, ideally the boundary data is such that all the time derivatives vanish initially. A condition that is only approximatively realised in practice.
\subsection{Quasi-normal amplitudes}\noindent

\noindent
Quasi-normal modes are the complex $s_n-$values for which $\phi_n(\rho)$ is a {\em regular} solution to the homogeneous equation
\beq
{\boldsymbol \alpha} [\phi_n] + s_n {\boldsymbol \beta} [\phi_n]  = 0 \label{eq:QNMEq}.
\eeq
This equation is solved in {\texttt Mathematica} within the framework of the ``Generalised Eigenvalue problem" (see appendix \ref{sec:SpecMeth_QNM} for further details).

As discussed in \cite{Ansorg:2016ztf}, the function $\bar{U}(\rho;s)$ has poles on the quasi-normal modes. We introduce the decomposition
\beq
\bar{U}(\rho;s) = \frac{\bar{V}(\rho;s)}{s-s_n} + \bar{W}(\rho;s).
\eeq
Besides, we identify the differential operator on the l.h.s of equation~\eqref{eq:LapTransEq} as ${\mathbf A} (s)= {\boldsymbol \alpha} + s {\boldsymbol \beta}$ and rewrite it in the form
\beq
{\mathbf A} (s) = {\mathbf A} (s_n) + (s-s_n){\boldsymbol \beta}.
\eeq
Then, equation~\eqref{eq:LapTransEq} in the limit $s \rightarrow s_n$ gives as regularity condition
\beq
\label{eq:AmpliEq_ZeroOrder}
{\mathbf A} (s_n) [ \bar{V} ] = 0 \Rightarrow \bar{V} = \eta_n \, \phi_n(\rho).
\eeq
At second order, we obtain
\beq
\label{eq:Amplitude}
{\mathbf A} (s_n) [ \bar{W} ] + \eta_n \, {\boldsymbol \beta}[\phi_n] = {\boldsymbol \beta} [U_{\rm in}]- \bar{S}.
\eeq
 equation~\eqref{eq:Amplitude} is to be solved simultaneously for $\bar{W}(\rho)$ and $\eta_n$ with the normalization condition $\bar{W}(\rho_0)=\bar{W}_0$ (for further details see appendix \ref{sec:SpecMeth_QNM_Amplitudes}). \\

Then, the standard recipe to obtain the spectral representation of the solution $U(v, \rho)$ is as follows: 
\begin{enumerate}
\item one expresses the time-dependent solution $U(v,\rho)$ as the inverse Laplace-transformation;
\item the integration path of the inverse Laplace-transformation in the complex $s-$plane is deformed in order to include the QNMs (see \cite{Ansorg:2016ztf} for further details). By doing so, we obtain
\beq
\label{eq:SpecDecom_GenSol}
U(v,\rho) = \sum_{n=0}^{\infty} \eta_n \phi_n(\rho) \e^{s_n v} + \int_{\cal C} \dd s\, \hat{U}(\rho;s) \e^{s v} ,
\eeq
with the last term corresponding to the integral along the half-circle ${\cal C}$ as $|s|\rightarrow \infty$; 

\item for asymptotically flat spacetimes, it is argued in~\cite{Ansorg:2016ztf} that the validity of \eqref{eq:SpecDecom_GenSol} depends on the behaviour of $|\eta_n \phi_n(\rho)|$ for large $n$. It is shown that generically $|\eta_n \phi_n(\rho)| \sim \e^{-\tau \text{Re}(s_n)}$. For $v< \tau$, the function $\hat{U}(\rho;s)$ diverges exponentially as $|s|\rightarrow \infty$. On the other hand, for $v>\tau$, the contribution from the integral $\int_{\cal C}$ is zero and we obtain the desired spectral representation.
\end{enumerate}
\subsection{Numerical evidences for the recipe in AdS-Schwarzschild}
\begin{figure}[h!] 
\centering
\includegraphics[width=6.9cm]{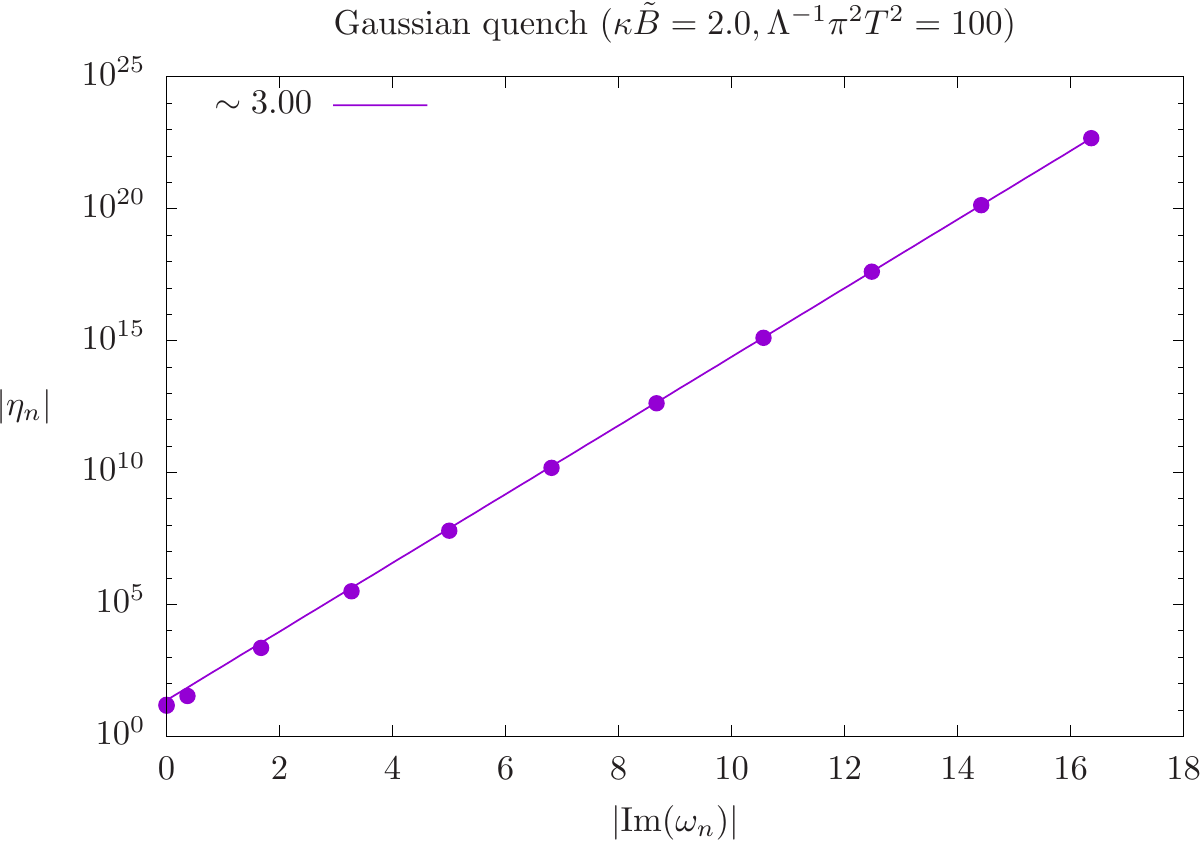}
\hspace{1cm}
\includegraphics[width=6.9cm]{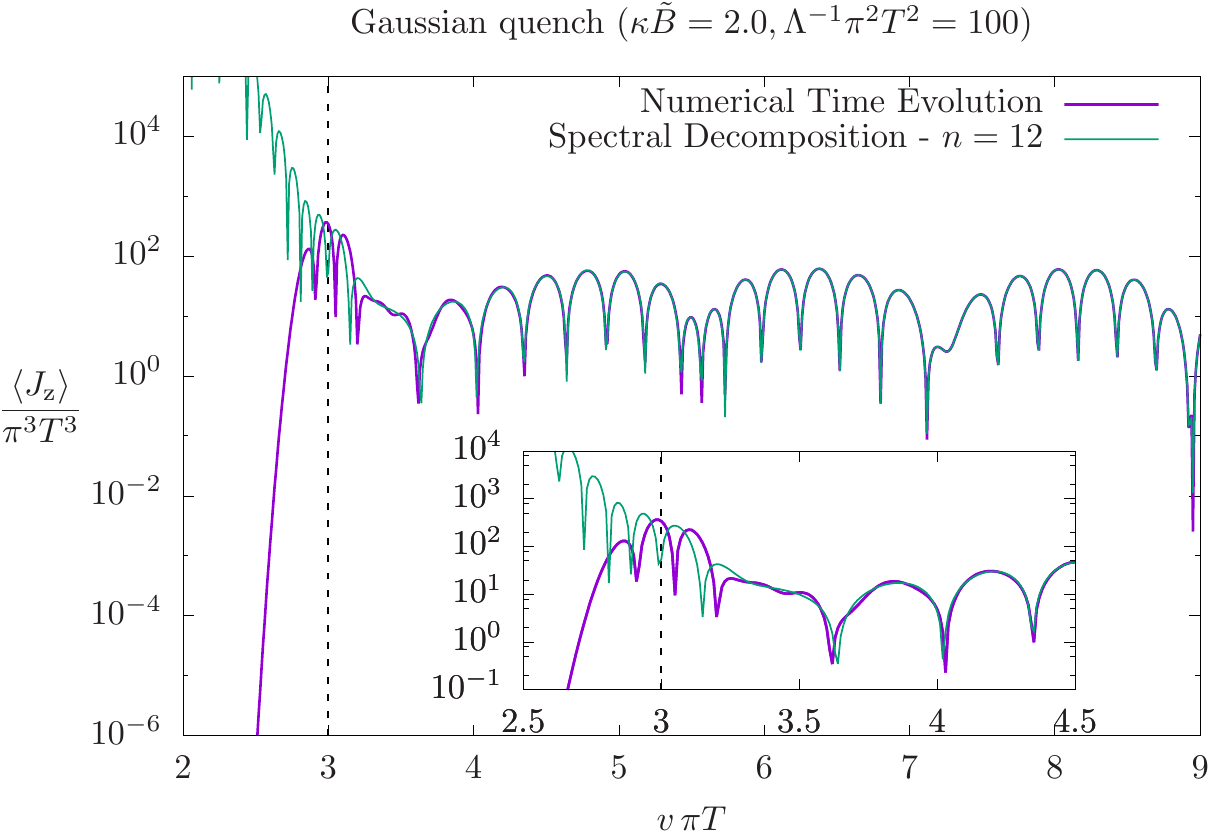}
\caption{\label{fig:123} Left: Logarithmic plot for the amplitudes against $\text{Im}[\omega]$ for the first twelve QNMs. Right: Comparison of the spectral analysis (green) for twelve QNMs to the full current (purple) obtained by solving equation \eqref{eq:vzeq} for a generic Gaussian quench. The chosen parameters  yield $\tau=3$. The zoomed image shows times very close to $\tau$ where the convergence is not yet achieved. }
\end{figure}

\noindent
\noindent
Here, we present numerical evidence showing that the recipe discussed above may also work for asymptotically AdS spacetimes. Using \eqref{eq:Current} we identify\footnote{As detailed in appendix \ref{sec:SpecMeth_QNM_Amplitudes}, we can always normalise $\phi_n(0) = 1$.} \begin{equation}A_n =-2\, \eta_n \phi_n(0).
\end{equation}
First, for the quenches to be presented in~\eqref{eq:Quenches}, we calculate the quasi-normal amplitudes $A_n$ using equation~\eqref{eq:Amplitude} and we observe that they indeed behave as
\begin{equation}\label{eq:tautau}
|A_n|\sim \e^{-\tau\text{Im}[\omega_n]}\,.
\end{equation}
Here $\omega_n$ are the Fourier frequencies. We expect\footnote{It is possible to fine tune the initial data to excite only one (or several) QNMs. In this cases the argument is not valid.} this behaviour to be valid for any quench which approaches a stationary configuration as $v \rightarrow \infty$. In particular, we show in the left panel of figure~\ref{fig:123} an example for the Gaussian quench with $\kappa B=2$ and $\Lambda =0.1$, for which we read the time $\tau = 3.0$.

Second, we check that the sum in \eqref{eq:spectraldec} converges for $v>\tau$, based on 
\begin{equation}
|A_n|\e^{-\im\omega_n v} \Big|_{n\rightarrow\infty} \sim \e^{-\text{Im}[\omega_n](\tau-v)}\e^{-\im\text{Re}[\omega_n] v}\,.
\end{equation}%
In figure \ref{fig:123} we also show the comparison of the spectral analysis to the numerical data obtained by solving the PDE for this specific quench. As mentioned before, here we obtain $\tau=3.0$ and the time evolution based on the spectral decomposition matches the numerical evolution accordingly. 

In practice, since we can only take a finite number of quasi-normal modes into account, we can distinguish three regimes in the plot. The first is $v<\tau$ where the spectral decomposition diverges and does not resemble the current response at all. There is a second, intermediate, regime $\tau< v\lesssim 4$ where the fit is not completely accurate. In order to obtain a perfect fit in this region, we should also take into account higher QNMs. Finally for $v \gtrsim 4$ the spectral decomposition fits perfectly the numerical data.

The regime $v < \tau$ requires further investigations. It is still not clear to us whether the inclusion of all higher QNMs and possibly an analytic continuation can cure the deviation between the numerical time evolution and the spectral decomposition observed in figure \ref{fig:123}. We leave further investigation in this topic for future work. 

To summarise, $\tau$ defines the time from which an explicit solving of the time evolution can be substituted by an analysis based on the spectral decomposition i.e. on the quasi-normal modes of the system. It depends not only on the system under consideration but on the concrete initial and boundary conditions. Moreover, there is in principle no bound on the value of this quantity. This means that for specific systems or initial/boundary conditions it might be possible to get $\tau<0$. 

In \cite{Szpak:2004sf}, it is argued that $\tau$ can be understood as the time it takes for (a compact support) initial data to propagate to a point whose lightcone contains all the initial data information. So far, we have not found a similar intuition in AdS. It is worth mentioning that \cite{Szpak:2004sf} assumes that the initial data is a function of compact support and does not consider the possibility of a boundary with ``ingoing'' data.

In the following, we investigate the dependence of $\tau$ on the width of the quenches and on the effective anomaly-magnetic field parameter $\kappa B$. With this we would like to explore the possibility of considering $\tau$ as a well defined notion of ``initial time'' for linear (or linearised) systems.

\section{Initial time}\label{sec:initial}\noindent
 In the previous section we reviewed how to compute the residues of the QNMs of the system and discussed the mutual growth rate $\tau$. As emphasised before, this time scale is fixed not only by the system under consideration but by the initial and boundary data as well. In this section we study the dependence of $\tau$ on the different quenches and on $\kappa B$ and we explore the possibility of identifying it with the out of equilibrium or ``initial response'' time of the system.

Concretely we have computed $\tau$ as a function of the anomaly parameter $\kappa B$ for quenches of the form 
\begin{equation}
\label{eq:Quenches}
V_0(v) = \left\{
\begin{array}{ccc}
\text{exp} \left \lbrace-(v-v_i)^2\Lambda\right\rbrace & & {\rm (Gaussian \, quench)} \\
\left[ 1+ \tanh\left \lbrace (v-v_i)\Lambda\right\rbrace\right]/2\,  & & {\rm (tanh \,  quench)},
\end{array}\right.
\end{equation}%
%
\begin{figure}[t!] 
\centering
\includegraphics[width=6.9cm]{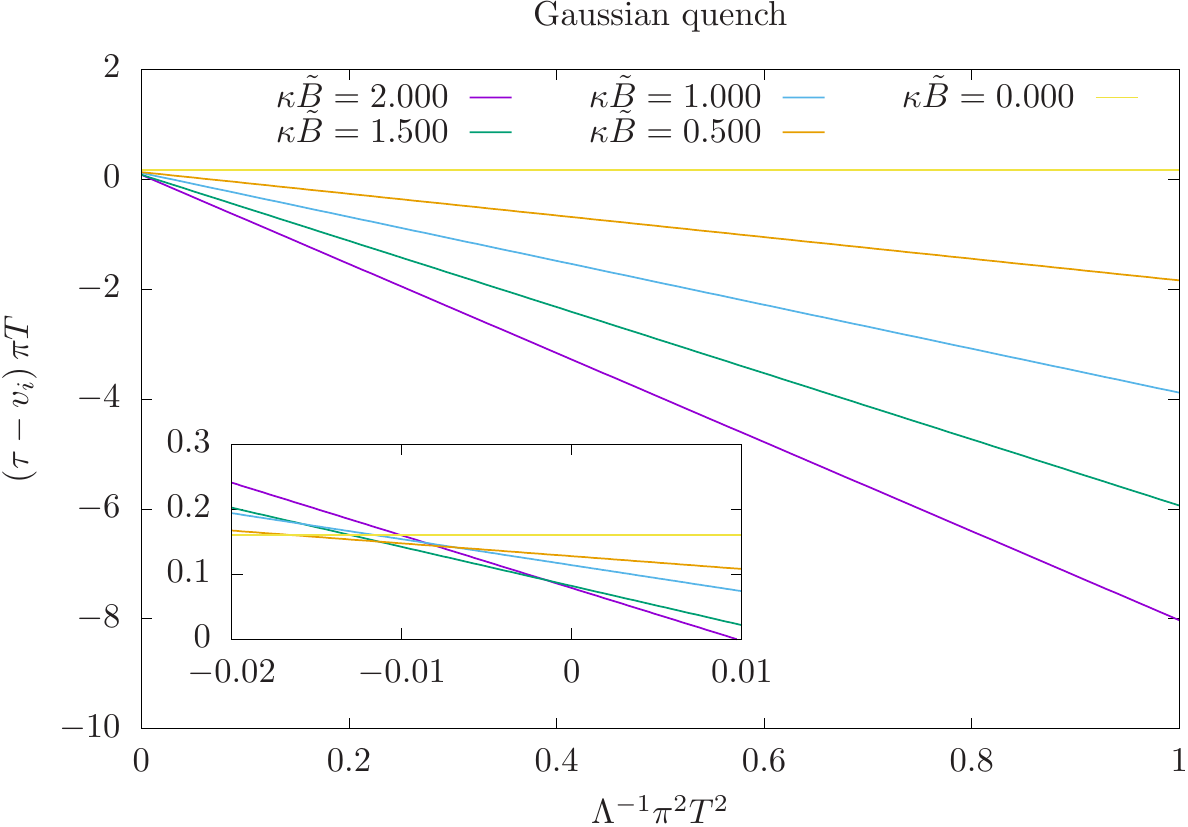}
\hspace{1cm}
\includegraphics[width=6.9cm]{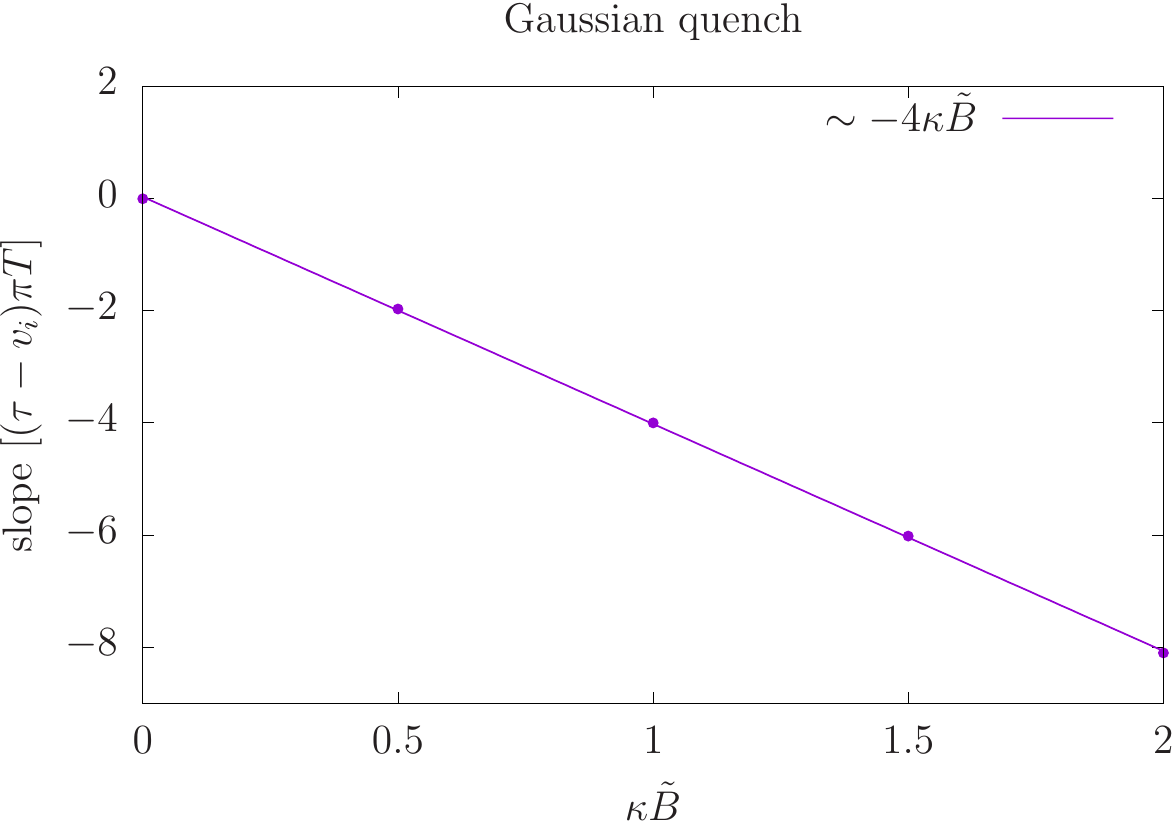}
\caption{\label{fig:tau1}Left: Distance of $\tau$ to the centre of the Gaussian quench against its width for several values of $\kappa \tilde{B}$. From now on we use the rescaled magnetic field $\tilde B \equiv B/(\pi T)^2$ in the plots. In the zoom the ``very fast'' quench region is shown. Right: Slopes of the lines in the l.h.s. against $\kappa \tilde{B}$. The data fits $\sim -4\kappa \tilde{B}$.}
\end{figure}
where $v_i$ fixes the ``centre of the source", and of $\Lambda$, which is inverse to the width of the signal. This quantity should be though of as the abruptness of the quench, so that $\Lambda\gg1$ implies a very fast quench. Let us remark that the main qualitative difference between the two sources is that Gaussian sources introduce no final net axial charge in the system while hyperbolic tangents do. This can be easily understood by considering that the total axial charge introduced in the system goes like\footnote{ Equation \eqref{eq:totalcharge} makes it clear that the total charge generated is of topological nature.} 
\begin{equation}\label{eq:totalcharge}
\rho_5\sim \int \dd v \, E\cdot B = B\, V_0(v)\Big|^{v_f}_{v_i}\,.
\end{equation}%
%
\begin{figure}[t!] 
\centering
\includegraphics[width=8.cm]{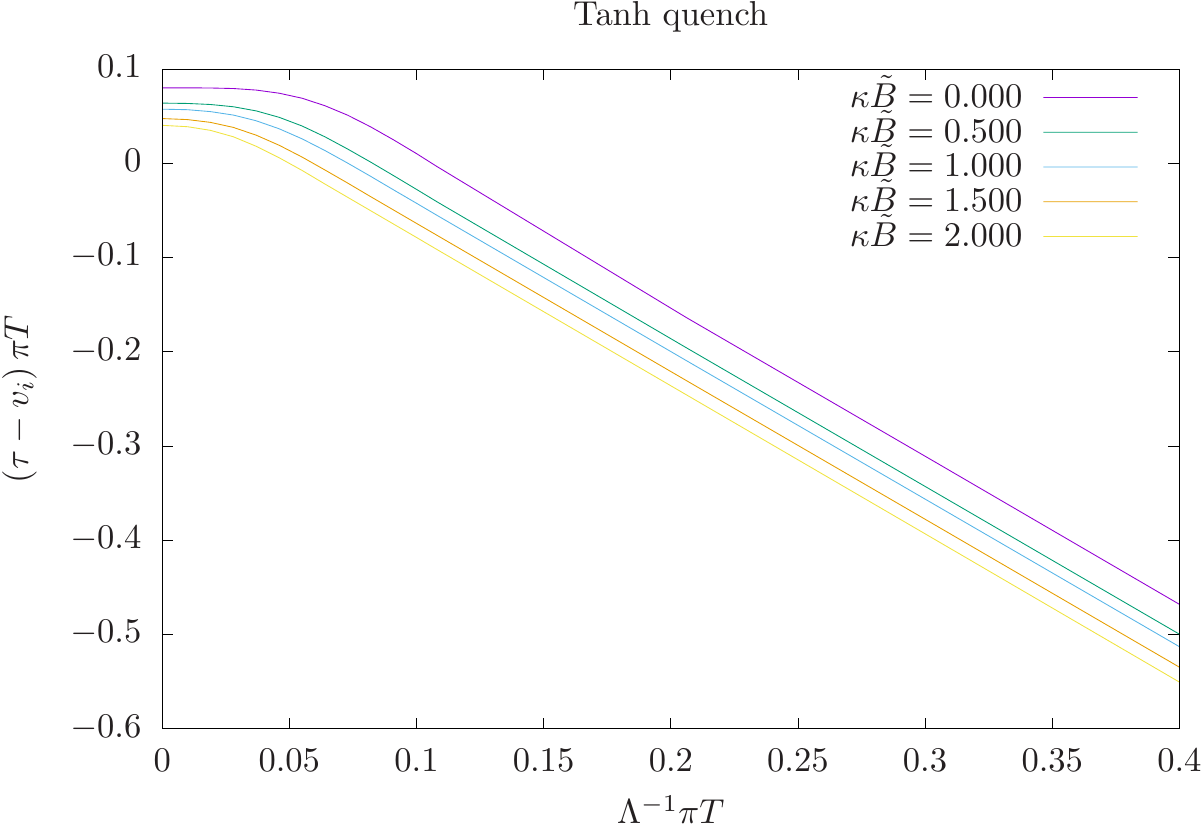}
\caption{\label{fig:tautanh} Distance of $\tau$ to the centre of the tanh quench against $\Lambda^{-1}$ for several values of $\kappa \tilde{B}$. Note that $\tau$ deviates from the linear behaviour for fast quenches i.e. $\Lambda^{-1}$ small. In the linear regime the slope is independent of $\kappa \tilde{B}$.}
\end{figure}
As in previous studies of holographic quenches  \cite{Buchel:2013lla}, we expect the quench scale $\Lambda$ to play an important role in the response of the system. For non-compact support functions like the ones we are considering, it is necessary to set some reference time to compare $\tau$ with. The most natural quantity to consider in our case is the temporal distance to the centre of the source: $\tau-v_i$. In figures \ref{fig:tau1} and \ref{fig:tautanh} we show the dependence of $\tau-v_i$ on $\Lambda$ for several values of $\kappa B$ for Gaussian and tanh quenches respectively. 

For Gaussian quenches (see figure~\ref{fig:tau1}) we find a linear dependence on $\Lambda^{-1}$. In particular, for $\kappa B=0$ we observe no dependence on this quench parameter. In the r.h.s. of figure \ref{fig:tau1} we show the dependence of the slope $(\tau-v_i)/\Lambda^{-1}$ with $\kappa B$. As shown in the figure the data fits to a line with slope $-4\kappa B$. The fact that this slope is negative means that for the same quench, the initial time is smaller the higher $\kappa B$. Hence, the system is closer to the adiabatic response the stronger the anomaly term. In other words, the relaxation time of the system is smaller for larger $\kappa B$.

For tanh quenches (see figure~\ref{fig:tautanh}) we find again a linear dependence on $\Lambda^{-1}$ for small $\Lambda$ ($\Lambda\lesssim 10$). In this case, contrary to what we observed in the Gaussian quench, $\kappa B$ only shifts the lines and does not affect the slope. For larger $\Lambda$, the time scale $\tau - v_i$ shows a non-linear dependence on $\Lambda^{-1}$. Although qualitatively different, the behaviour of $\tau$ for tanh quenches is still compatible with the intuition of a smaller relaxation times the higher $\kappa B$. 
 
Summarising, in all cases the data fits to a linear dependence on $\Lambda^{-1}$ in the regime of small enough $\Lambda$. That $\tau-v_i$ is generically negative and increases for faster quenches seems natural, since for very slow quenches one gets closer to the adiabatic regime and one expects to find $\tau-v_i\rightarrow-\infty$. 
 
By looking at a different quantity, obtained within the initial time $v<\tau$, we will confirm this intuition in next section.

 \subsection{Universality in fast quenches}\label{sec:adiabatic}\noindent

\begin{figure}[t!] 
\centering
\includegraphics[width=6.9cm]{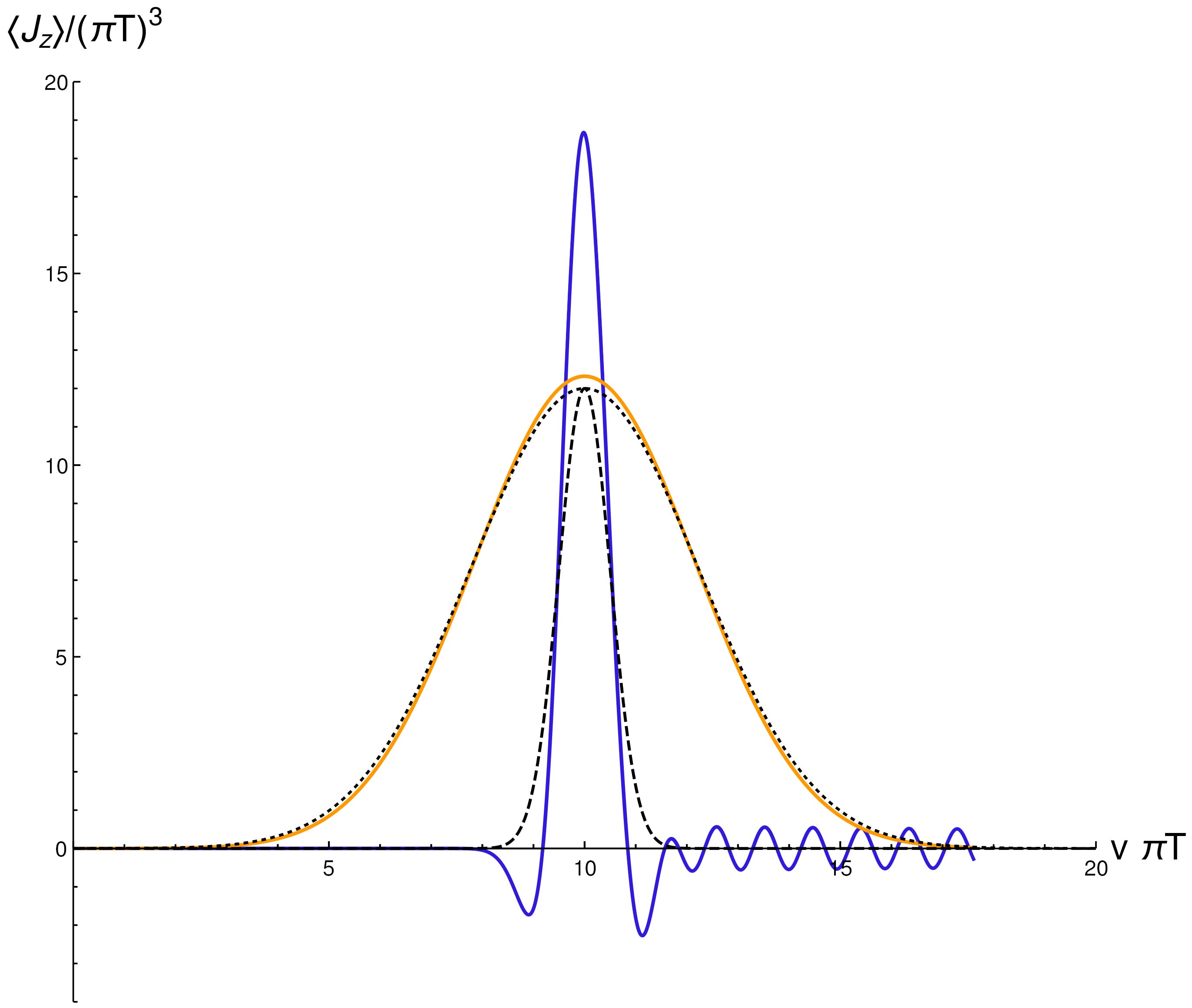}
\hspace{1cm}
\includegraphics[width=6.9cm]{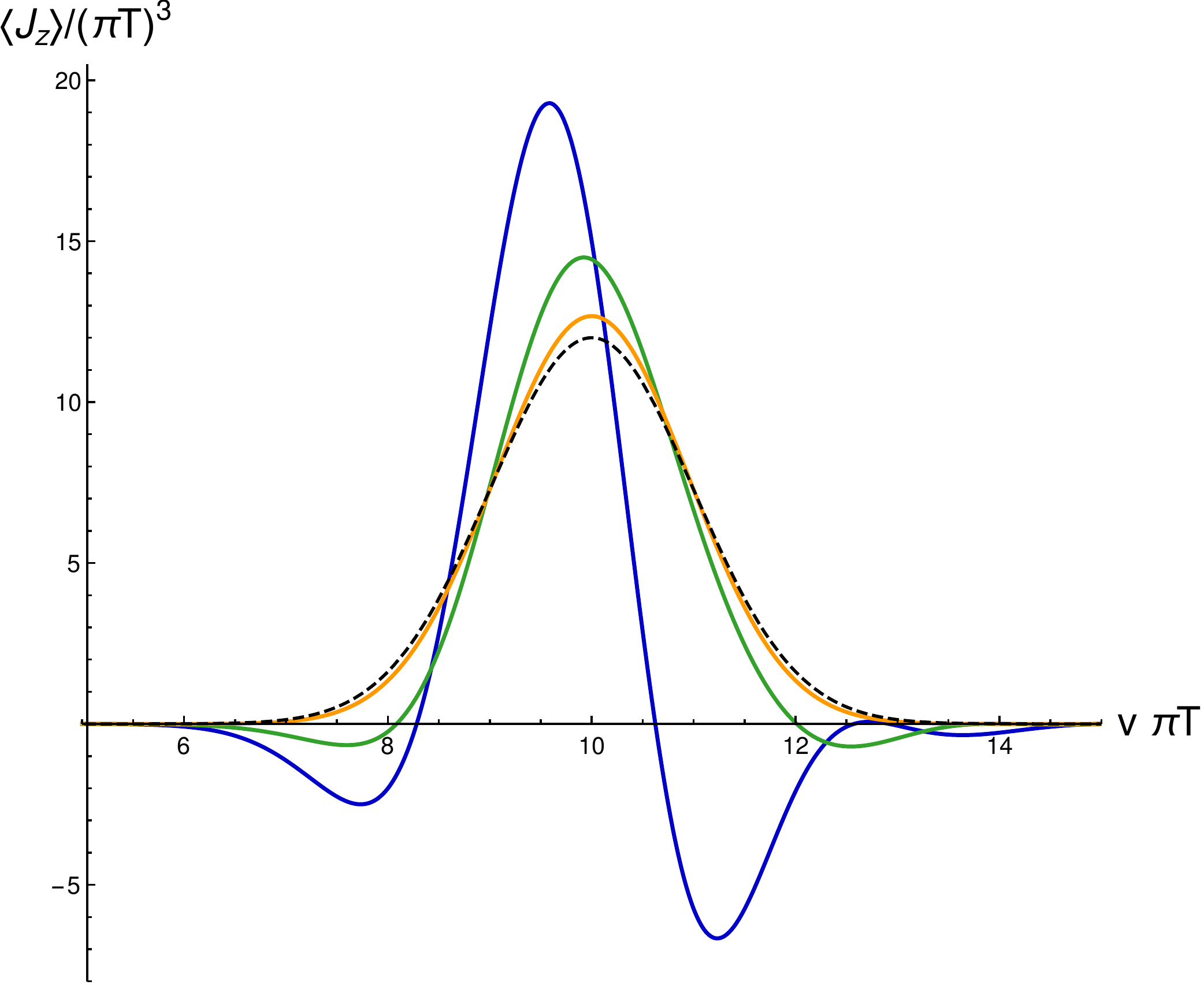}
\caption{\label{fig:adiabvsnoadiab}Left: Current against time for Gaussian quenches with same height and different widths $\Lambda/(\pi T)^2=0.1$ ``slow'' (orange) and $\Lambda/(\pi T)^2=2$ ``fast'' (blue) at fixed magnetic field $\kappa \tilde{B}=1$. Dashed back lines show the adiabatic response given by equation \eqref{eq:adiabatic}. Right: Current against time for fixed Gaussian source with $\Lambda/(\pi T)^2 =0.5$ and several values of the magnetic field $\kappa \tilde{B}=0.1,0.5,4$ (blue, green, orange). For an easier comparison, the currents have been rescaled such that the corresponding adiabatic response (black, dashed) is the same.}
\end{figure}
\noindent
 In the previous subsection we have seen how $\tau$ depends on $\Lambda$ and $\kappa B$. As already argued this provides us with some notion of ``initial'' response of the system. In this subsection we investigate the initial response regime defined as $v<\tau$. Let us recall that for this part of the response one has to, by definition, explicitly integrate equation~\eqref{eq:DynEq_TimeDom} in time.
 
 Our basic motivation comes from the studies of quantum quenches in \cite{Buchel:2013lla, Das:2014hqa}. In these papers a universal response was found for certain kind of smooth, fast enough quenches in generic CFT's. Concretely the authors of \cite{Buchel:2013lla} looked at a free scalar holographic model. Among others, it was shown that the time $\tau_\text{ex}$, at which the operator deviates 5\% from the adiabatic response, follows a simple universal law for fast quenches. We perform an analogous analysis in our system. 
 
 Before we proceed to explain the details, some comments are in order.
 A first observation is that our system is linear, in contrast with that of \cite{Buchel:2013lla}. As we will see this does not prevent the appearance of an universal behaviour for fast quenches. In addition to this, we emphasise that our main interest is to determine how the anomaly affects the universal behaviour. Therefore we focus on the dependence of the universality on the value of our anomaly/magnetic parameter $\kappa B$. Moreover, we focus on Gaussian sources henceforth.  
\begin{figure}[t!] 
\centering
\includegraphics[width=6.9cm]{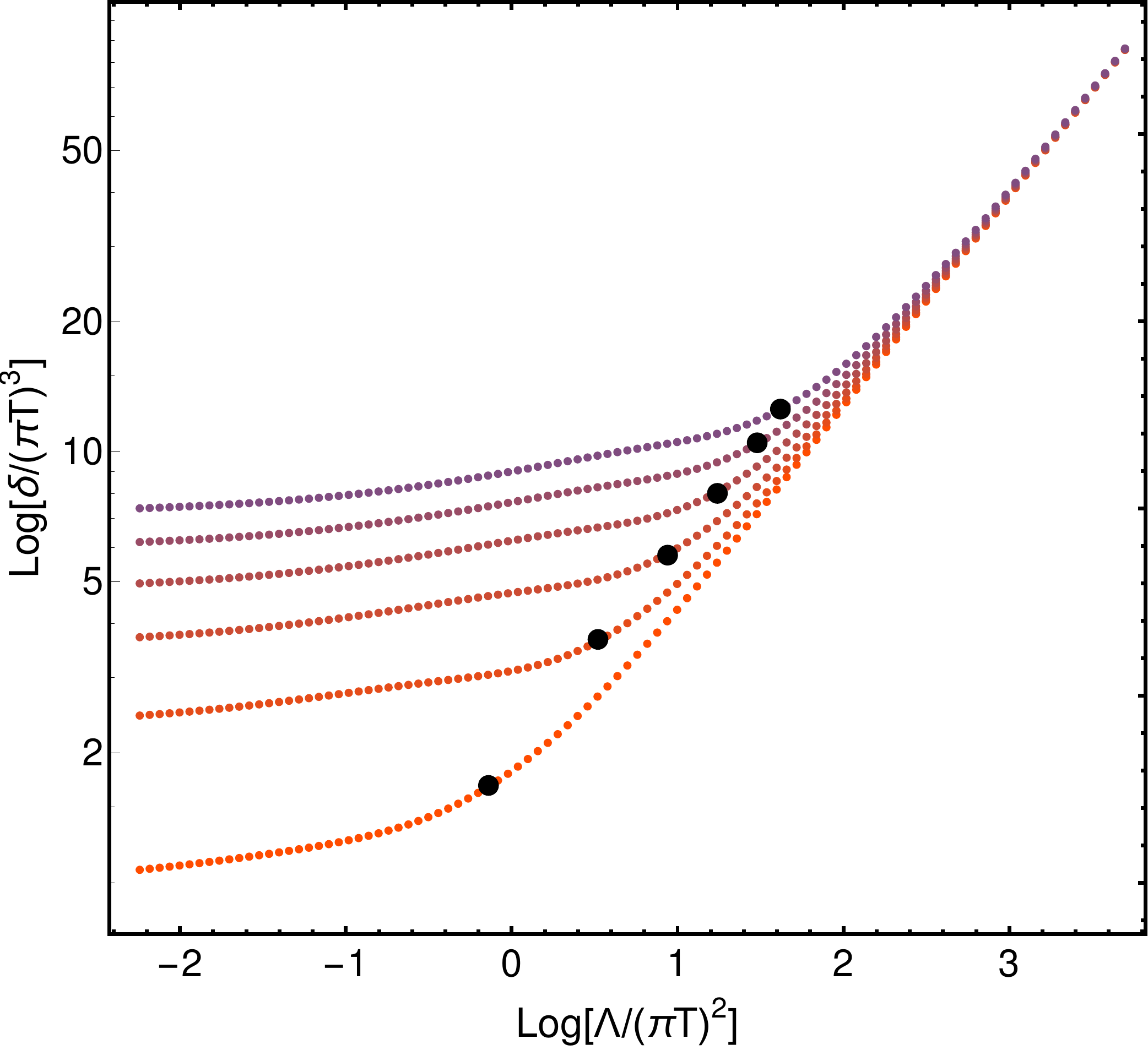}
\hspace{1cm}
\includegraphics[width=6.9cm]{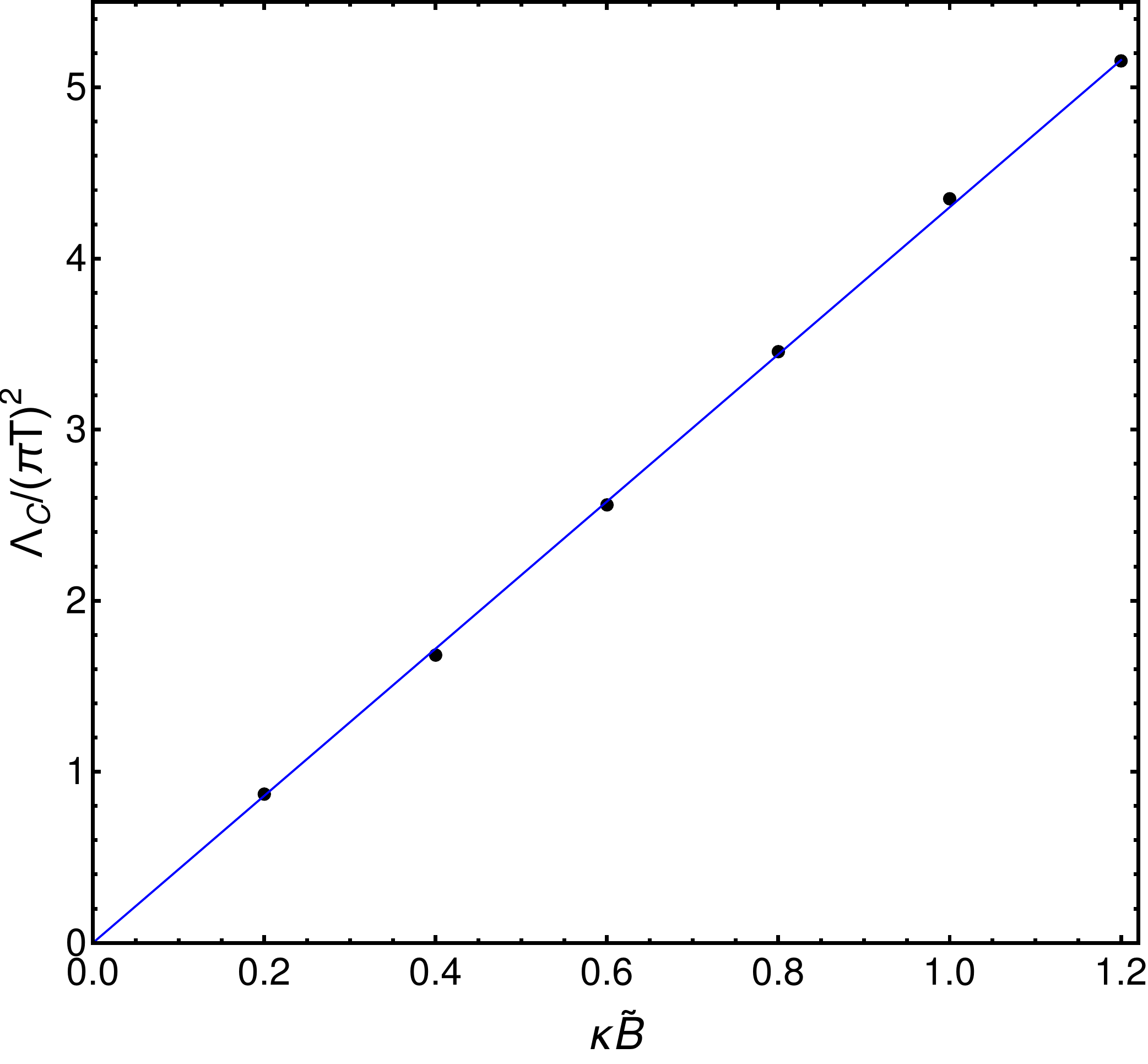}
\caption{\label{fig:max-min}Left: Logarithmic plot of $\delta$ (see equation \eqref{eq:delta}) against the abruptness of the quench $\Lambda$ for several values of $\kappa \tilde{B} =0.2-1.2$ (red-blue). For high enough $\Lambda$ all cases converge to a straight line with slope 1.120. Black points highlight the critical $\Lambda_C$ as defined in 
the text. Right: Critical inverse width $\Lambda_C$ against $\kappa \tilde{B}$ (black) which can be fitted by a straight line (blue).}
\end{figure}
To give a qualitative idea of the system response, we show the generic behaviour of the current in figure \ref{fig:adiabvsnoadiab}. In the l.h.s.~we fix $\kappa B$ and plot the current for two qualitatively different values of the width $\Lambda$ of Gaussian.
  The qualitative difference between a ``fast'' quench (blue)  and a ``slow''  quench (orange) is apparent: in the slow case the current approaches the adiabatic behaviour, given by equation~\eqref{eq:adiabatic}, depicted in black. The fast quench shows a more complicated structure and deviates significantly from its corresponding adiabatic behaviour. In the r.h.s.~we fix the width and show the response for different values of the magnetic field. Again, one can observe a qualitative difference; the lower $\kappa B$ the bigger the deviation from the adiabatic response. This simple qualitative observations seem consistent with those made in the previous section. We would like now to see whether the system shows an universal behaviour for fast quenches. This will shed some light concerning the relaxation time. In order to characterise this we have found it useful to look at the following quantity
\begin{equation}\label{eq:delta}
\delta\equiv |J_z(v_1)|-|J_z(v_2)|,
\end{equation}%
with $J_z(v_1)$ and $J_z(v_2)$ the first minima/maxima of the current, respectively.
 We have computed $\delta$ as a function of $\Lambda$ for several values of the $\kappa B$. Our results are summarised in figure \ref{fig:max-min}. The plot shows that fast quenches indeed show an universal, $B$ independent, behaviour. Concretely we find 
 \begin{equation}
 \delta\sim \Lambda^{\alpha} \quad {\rm with} \quad \alpha = 1.118 \,,
 \end{equation}%
 and a fitting error for $\alpha$ of the order $10^{-3}$. As one can observe in figure~\ref{fig:max-min}, as we increase $\kappa B$, we need faster quenches, i.e.~higher $\Lambda$, to get to the universal regime.  Despite of the smoothness of the transition, it is possible to define a critical $\Lambda$ using an interpolating function for the (logarithmic) data. It is natural to identify $\Lambda_C$ with the absolute maximum of the second derivative of the interpolating function. We get same results using either splines or Hermite polynomials as interpolants and for several interpolation orders.  In the left panel of figure \ref{fig:max-min} we have exaggerated the corresponding points. In the right panel we show the values of $\Lambda_C$ against $\kappa B$. By fitting this data to a curve of the form $a+bx^c$ we find $a=-0.08$, $b=4.33$ and $c=0.96$ with fitting errors 0.19, 0.19, 0.07 respectively. We conclude that the transition to the ``universal'', fast quench, regime is delayed linearly with $\kappa B$. This suggests that the effective relaxation time of the system is inversely proportional to $\kappa B$. This is in agreement with the $\tau$ behaviour for tanh quenches displayed in figure \ref{fig:tautanh} where we see that the nonlinear (``fast'') behaviour appears for higher $\Lambda$ the higher $\kappa B$. 
  
\section{Landau resonances}\label{sec:late}\noindent
\begin{figure}[t!] 
\centering
\includegraphics[width=230pt]{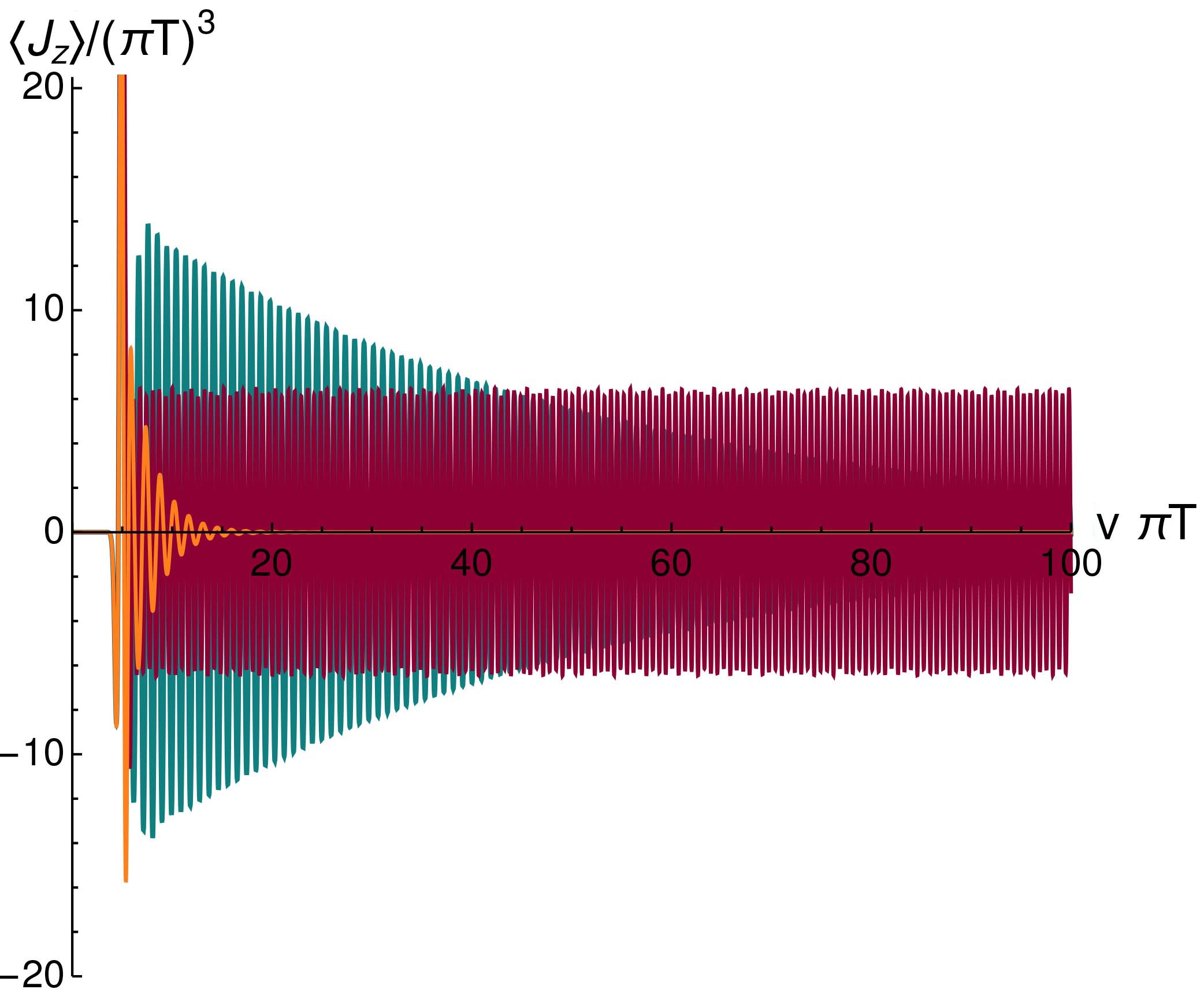}
\caption{\label{fig:latetime}Current against time for fixed Gaussian width $\Lambda/(\pi T)^2=6$ and several values of $\kappa \tilde{B}=0.5,1,2$ (orange, blue, red). Late time oscillations decay faster for smaller $\kappa \tilde{B}$.}
\end{figure}
\noindent
Let us now look at the response of the current for late times. Concretely, we consider times bigger than the transition time $\tau$ as defined in equation~\eqref{eq:tautau}. An interesting result in this regime is depicted in figure \ref{fig:latetime}. Here we show the behaviour of the current for a fixed Gaussian source and several values of the magnetic field. The decay rate of the late time oscillations decreases as we increase $\kappa B$. In particular, for $\kappa B\gtrsim 1$ this relaxation time is very small compared with all other scales in the system. This signals the existence of a QNM approaching the real axis for increasing $\kappa B$. We have computed the late time current for a variety of quenches (Gaussian, tanh, oscillatory) with analogous behaviour; the only qualitative difference we have found is the value around which the current oscillates and that is given by the total amount of axial charge induced in the system (see equation~\eqref{eq:totalcharge}).

There is, in fact, an analytic argument to see that indeed a QNM must approach the real axis with increasing $\kappa B$ by writing the perturbation equation for the relevant field component as a Schr\"odinger-type equation (see appendix \ref{sec:QNMapp} for further discussion).

Next we want to check whether this behaviour is indeed an anomaly driven phenomenon, at least for a class of holographic models. The main obstacle we face is that in the probe limit the magnetic field always appears in combination with the Chern-Simons coupling $\kappa$ in the equations. Therefore, in this limit it is not possible to disentangle anomaly from  magnetic field effects. Moreover, the probe approximation only makes sense for small values of the fields and we are now interested in the fate of the QNMs as we increase $B$. This motivates us to compute the QNM spectrum of the system including backreaction. Details on the computation can be found in appendix \ref{app:QNM}. Our main goal is to address the following questions: How do the lowest QNMs behave for increasing $B$? Is any mode crossing to the upper half plane $\text{Im}[\omega] >0$? What is the role of the axial anomaly in the dynamics of the QNMs?

\begin{figure}[t!] 
\centering
\includegraphics[width=6.9cm]{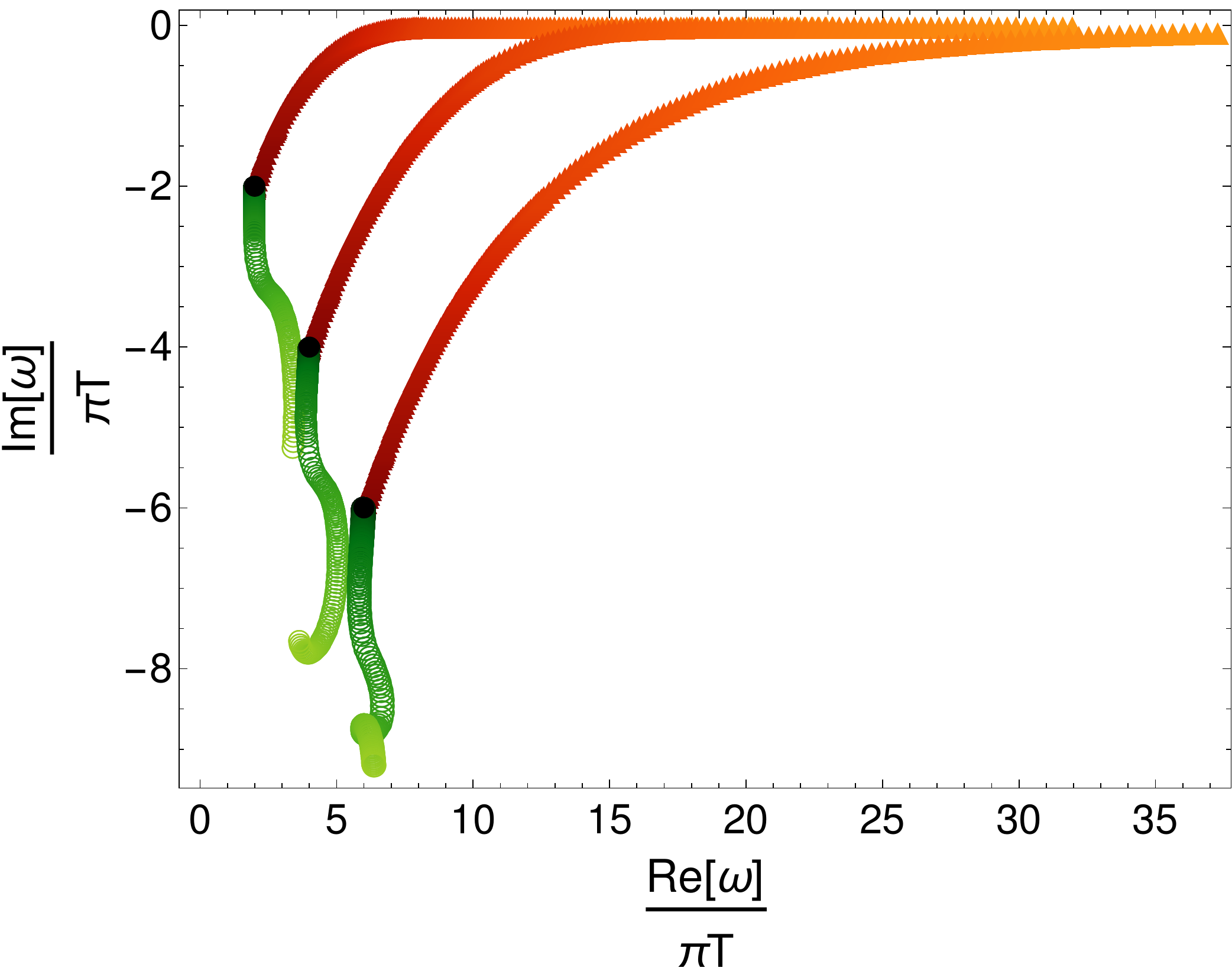}
\hspace{1cm}
\includegraphics[width=6.9cm]{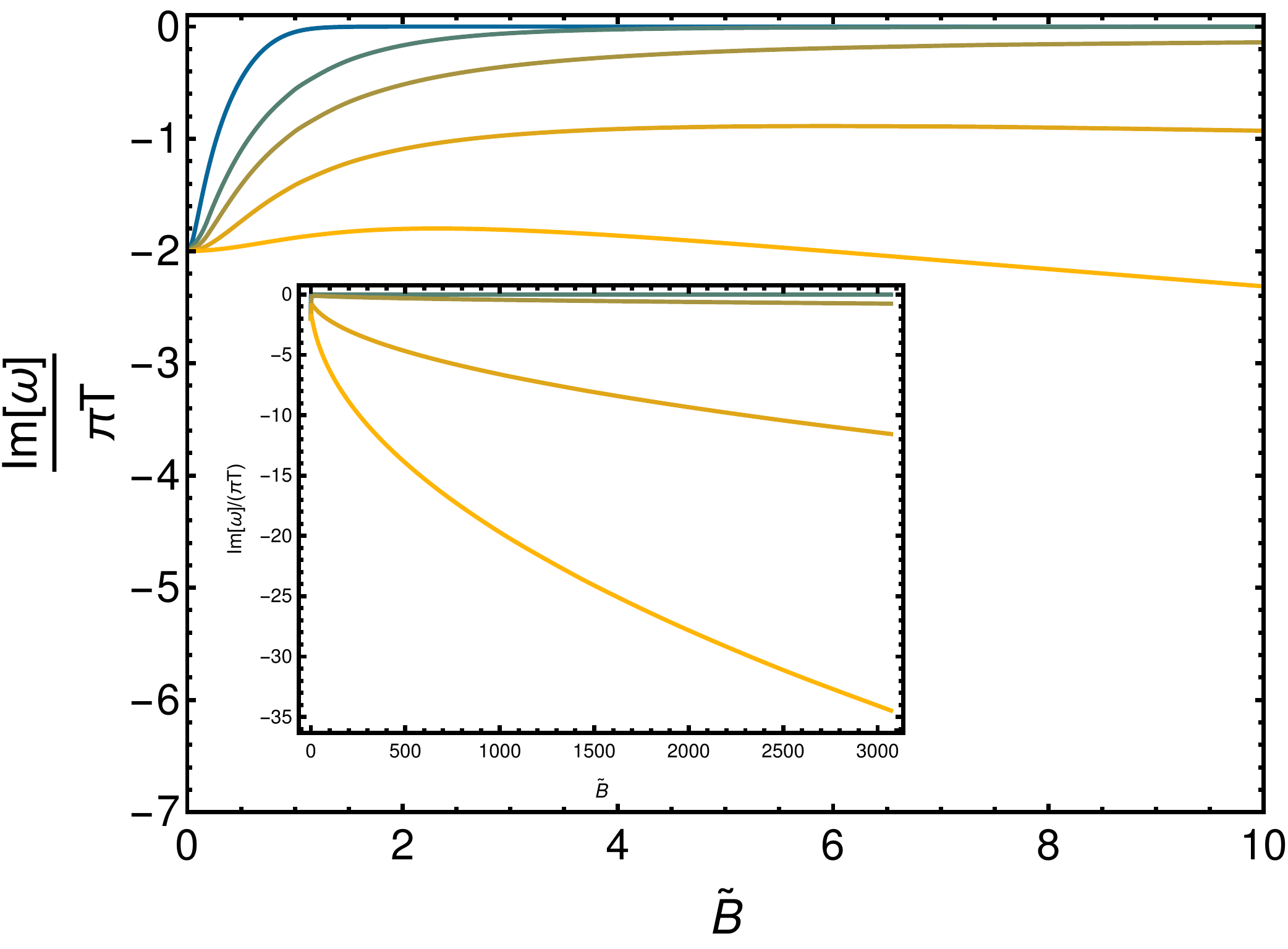}
\caption{\label{fig:imqnm}Left: Imaginary versus real part of the three lowest QNMs at zero frequency for  $\kappa=0$, $\tilde{B}=0-4$ (green-yellow) and $\kappa=1$, $\tilde{B}=0-4$  (red-yellow). Right: QNMs' imaginary part versus magnetic field for $\kappa\in\{ 0.1, 0.2, 0.34, 0.5, 1\}$ (yellow-blue).}
\end{figure}
\begin{figure}[t!] 
\centering
\includegraphics[width=6.9cm]{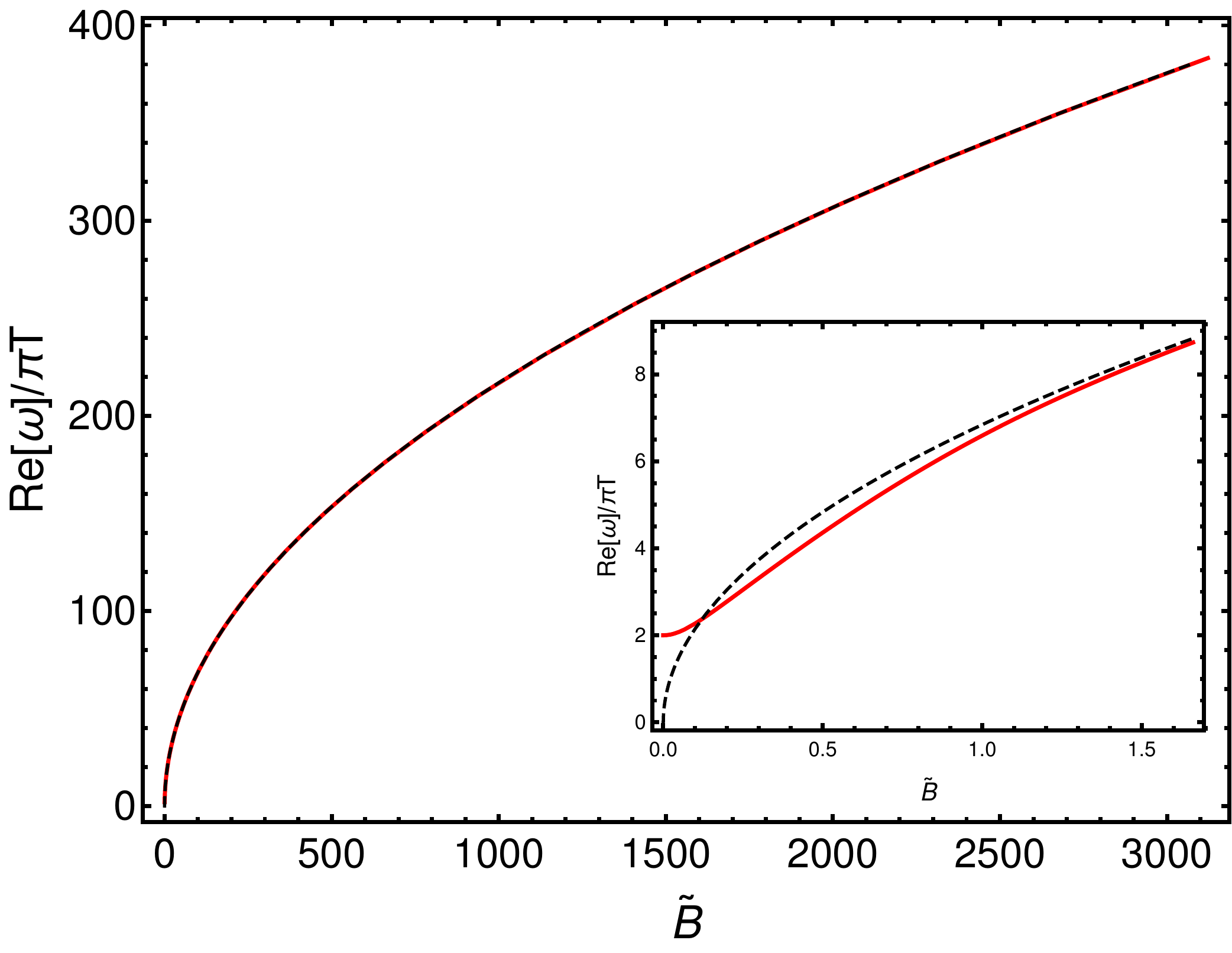}
\hspace{1cm}
\includegraphics[width=6.9cm]{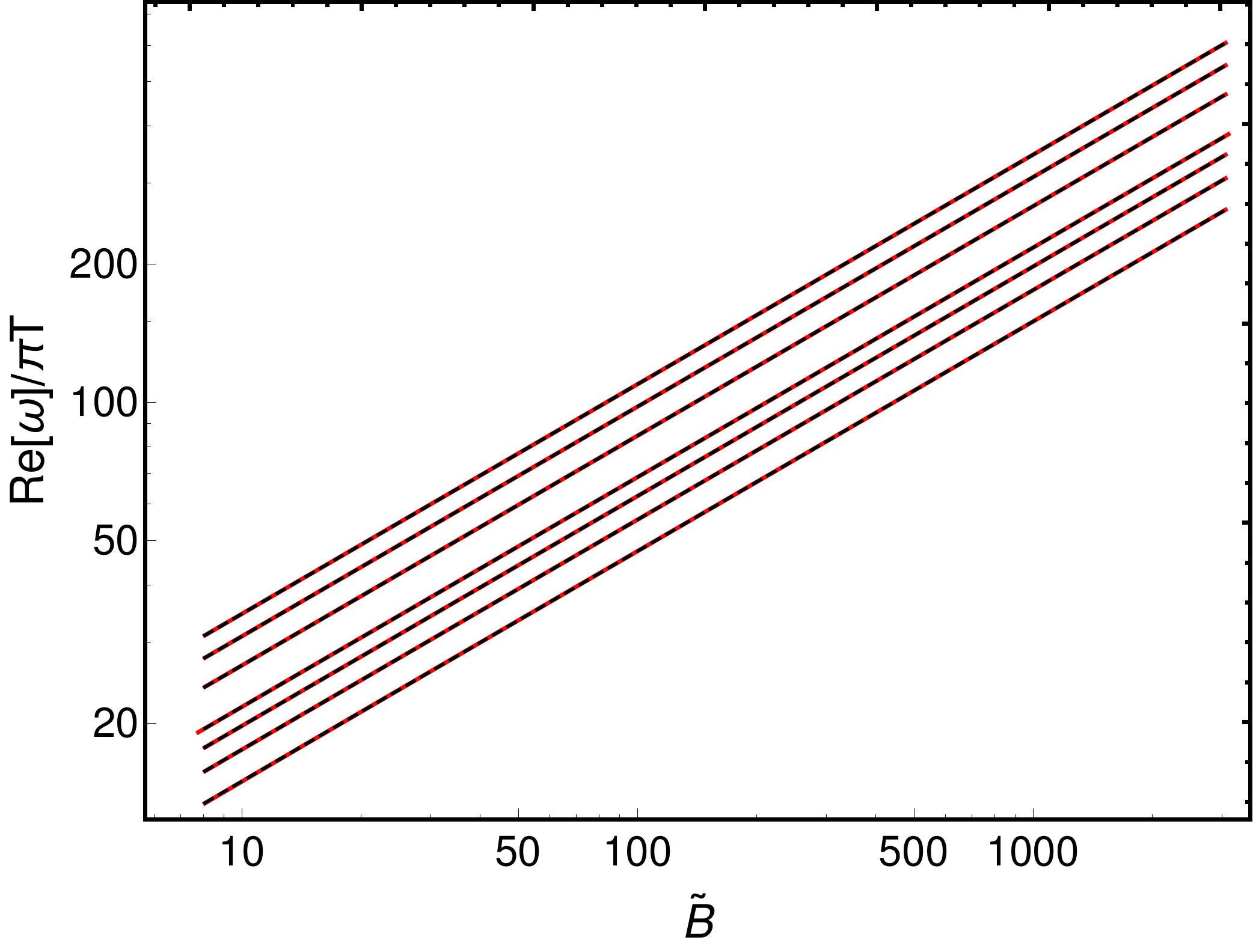}
\caption{\label{fig:reqnm}Left: Real part of the lowest QNM against magnetic field for $\kappa=1$ (red). For $\tilde{B}> 2 $ data fits $\sqrt{\tilde{B}}$ (dashed, black). Right: Double logarithmic plot of $\text{Re}[\omega]$ of the lowest QNM against $\tilde{B}$ for several values of $\kappa\in\{0.5, 0.67, 0.83, 1, 1.5, 2, 2.5 \}$  in the large magnetic field regime. Data (red) fits to $\sqrt{\tilde{B}}$ (dashed, black) in all cases.}
\end{figure}

We have looked at the dependence of the lowest QNMs on $\kappa$ and $B$ at zero momentum. The main message is summarized in the l.h.s. of figure \ref{fig:imqnm}. Here we show the behaviour of the three lowest QNMs as one increases the magnetic field for $\kappa=0$ and $\kappa=1$. The difference is apparent. For $\kappa=1$ the modes approach the real axis and the real part increases with B. Conversely, if we set $\kappa=0$ the modes approach the imaginary axis. There is an intermediate region, for approximately $0<\kappa\lesssim 1/2$. To illustrate this we show the fate of the lowest QNM for $\kappa\in\{ 0.1, 0.2, 0.34, 0.5, 1\}$ in the r.h.s. of figure \ref{fig:imqnm}. Here we see that in the low $\kappa$ regime the imaginary part of the frequency begins decreasing again once the magnetic field is large enough. We found similar results for higher QNMs. This implies that for low values of $\kappa$ no resonances are to be found, independently of the value of $B$. For $\kappa \gtrsim 1/2$ we find the mode approaching monotonically the real axis in the regime of B allowed by our numerics. For these values the QNMs approach the axis up to very small absolute values ($\text{Im}[\omega]\sim -10^{-6}$ for $\kappa=1/2$) of the imaginary part and resonances are found. In all cases the approach is faster the higher $\kappa$. We didn't find any mode crossing to the upper half plane for any of the values considered. 

In addition to this, it is clarifying to look at the behaviour of the real part of the frequency. By fitting it to a function of the form 
\begin{equation}\label{eq:fitx}
\text{Re}[\omega]=a+bB^c,
\end{equation} we find that far all non-zero values of $\kappa$ considered here, ranging from $\kappa=0.1$ to $\kappa=2.5$, the value of $c$ is compatible with $\sqrt{B}$ behaviour when $B$ is large enough. Concretely $c=0.4929$ (with a fitting error of $10^{-4}$) is the largest deviation from $c=1/2$ that we find. In the l.h.s. of figure \ref{fig:reqnm} we show the real part of the lowest QNM against magnetic field for $\kappa=1$. The zoom in the low $B$ region shows that the data deviate from the $\sqrt{B}$ behaviour at low values of $B$. In the r.h.s. of this figure we show the fit of this mode to $\sqrt{B}$ in the high $B$ regime for several, non-trivial, values of $\kappa$. This indicates that these resonances are a consequence of the presence of Landau levels in the system. At $\kappa=0$, however, the behaviour is qualitatively different, as one can already deduce from the l.h.s. of figure \ref{fig:imqnm}. In this case the real part of the frequency decreases with increasing $B$. This absence of Landau levels suggests that at $\kappa=0$ there are no fermions charged under the global symmetries induced by the bulk photons in the dual theory. Moreover the dependence of the imaginary part on $\kappa$ indicate that we can view the resonances as a footprint of the anomaly. 

Let us make a remark. In principle, for large values of $\kappa$ we cannot exclude the possibility of the modes going away from the real axis again for very large values of $B$. If this is the case, then the resonances would be restricted to a certain regime of the magnetic field. 

Next, we inquire the dependence of $b$ (see equation \eqref{eq:fitx}) on $\kappa$. For the data used in figure \ref{fig:reqnm} we find $b\sim\kappa^{0.510}$ with a fitting error of $10^{-3}$ for the exponent. Therefore, for high enough $\kappa$ and large enough magnetic field we conjecture that $\text{Re}[\omega] \sim \sqrt{\kappa B}$.\\   

\begin{figure}[t!] 
\centering
\includegraphics[width=6.9cm]{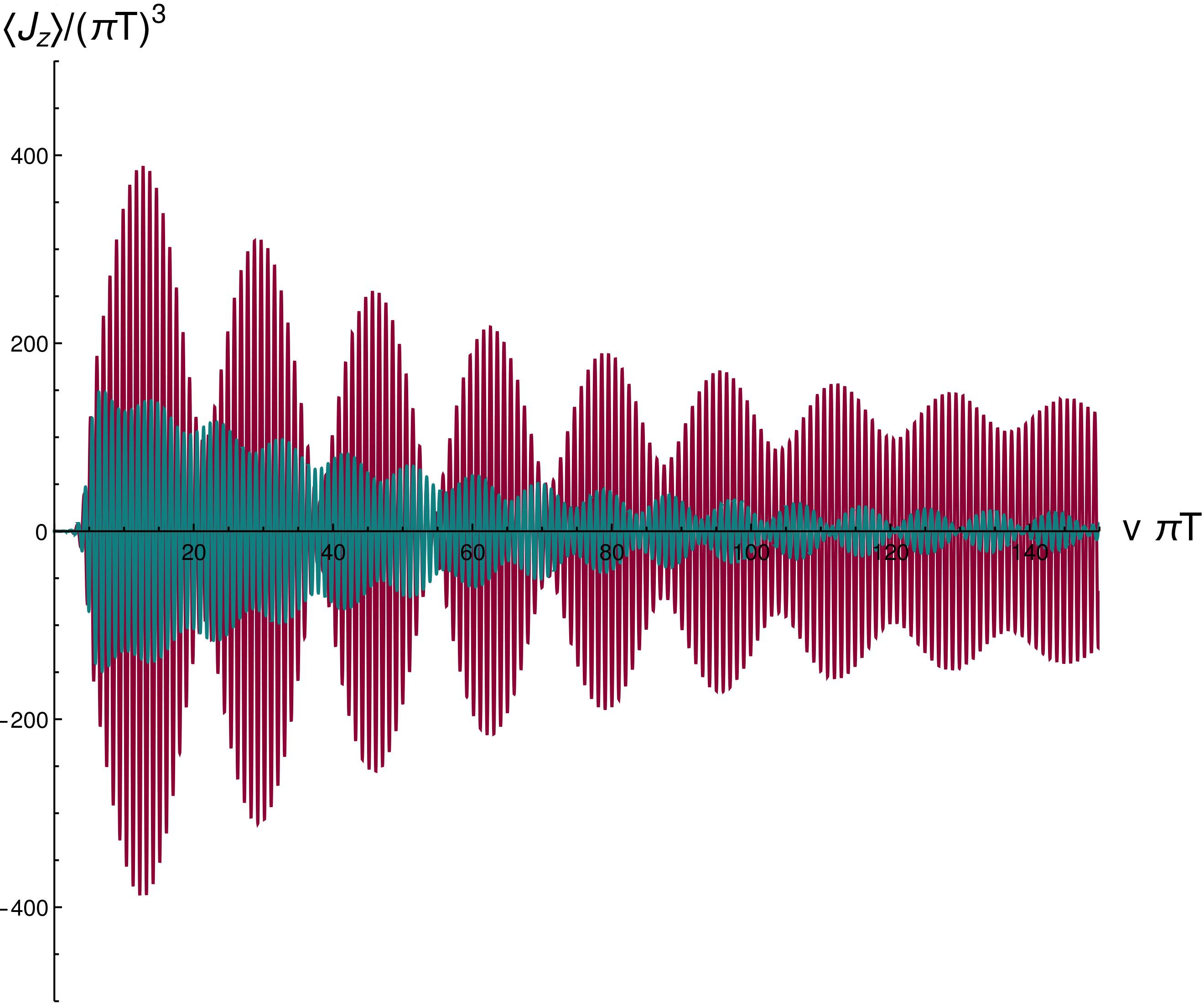}
\hspace{1cm}
\includegraphics[width=6.9cm]{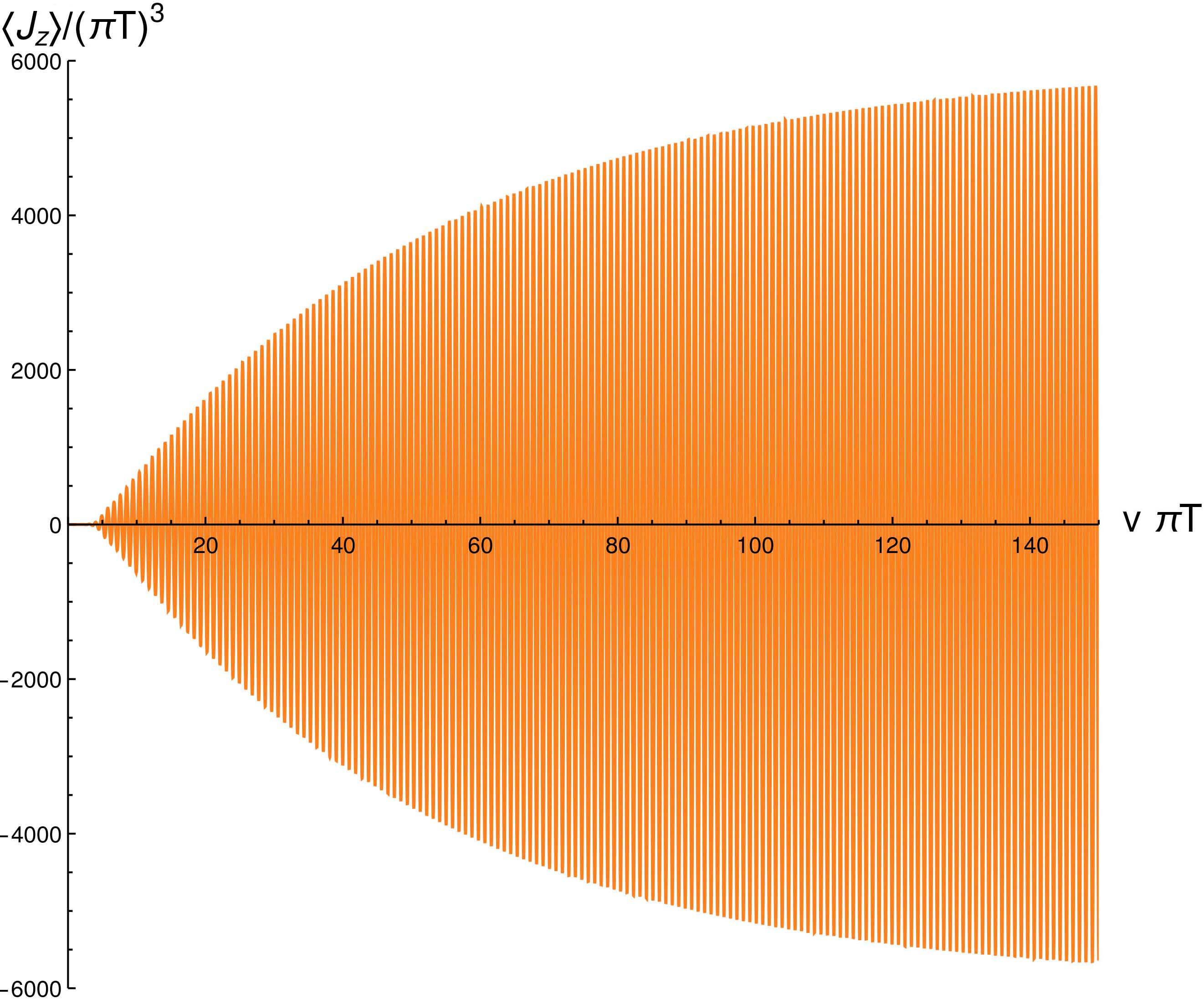}
\caption{\label{fig:resonance}Current vs time for an oscillatory source \eqref{eq:tanhsin} for  $\omega=0.95\omega_{\rm c}, 0.90\omega_{\rm c}$  (left: red, blue) and  $\omega=\omega_c$ (right: orange).}
\end{figure}

Finally, the slow decay rate of the current is not the only effect that ``almost normal'' modes have in the current. One can consider the possibility of directly exciting these modes by switching on an oscillatory source with appropriate frequency. In figure \ref{fig:resonance} we show the behaviour of the current within the probe limit for a source of the form
\begin{equation}\label{eq:tanhsin}
V_0(v)=(1+\tanh(v-v_i))\sin(\omega\,v)\, .
\end{equation}%
It corresponds to an oscillatory electric field with an amplitude envelope\footnote{The envelope plays no important role away from $v_i$ which is conveniently chosen to be small. One can check this by comparing the response of a damped oscillator to sources of the form $(\sin(\omega v))$ and $((\tanh(v-v_i)+1)\sin(\omega \,v))$.} chosen for numerical convenience.  As expected, the current response is qualitatively similar to that of a damped driven oscillator. For frequencies close to the resonant frequency $\omega_{\rm c}$ (fixed by $\kappa B$) the amplitude is dramatically enhanced and several characteristic frequencies appear. At the critical value of $\omega$ the amplitude is maximally amplified and it asymptotes a certain value. The higher $\kappa B$ the stronger the resonance. \\

\section{Discussion}\label{sec:discussion}\noindent
\noindent
In this paper we have considered quenches in an holographic system with a $U(1)\times U(1)$ anomaly. Concretely, we looked at the electric current generated parallel to a magnetic field known as chiral magnetic current and its response to a parallel time-dependent external electric field. Our motivations were twofold. On the one hand, as argued in the introduction, it is necessary to address the question of how anomalous transport behaves out of equilibrium and how it reacts to the dynamical evolution of the axial charge. On the other hand, we were interested in how the axial anomaly affects previous results \cite{Buchel:2013lla} regarding the universal response to fast quenches. 
As a first step, we have only considered the probe limit of the theory. The probe approximation limits the validity of the results to field configurations with a small stress-energy tensor compared to the temperature of the black hole. As a consequence, this approximation is only justified for sufficiently ``small" magnetic fields and the sufficiently ``slow"  quenches.
We have already made progress in the task of including backreaction into the system and will present our results in a follow up paper. However, we remark that one of our main results, the Landau resonances for large magnetic fields, were obtained with backreaction.  

In the probe limit, our setup yields in a linear hyperbolic PDE, which has to be solved as a initial/boundary value problem. We explicitly evolve this equation in time with a fully spectral method~\cite{Macedo:2014bfa}. Moreover, we also consider a different strategy to obtain the solution to the problem. Based on the discussion from~\cite{PhysRevD.45.2617, Nollert:1998ys,Ansorg:2016ztf} we read the so called quasi-normal amplitudes out of the initial/boundary data and express the solution as a spectral decomposition based on the quasi-normal modes. As discussed in~\cite{Ansorg:2016ztf,Szpak:2004sf} for asymptotically flat spacetimes, this analysis is expected to be valid for a time scale $v>\tau$, where $\tau$ is related to the growth rate  of the amplitudes. We observed that the same analysis works in our asymptotically AdS system. Moreover, based on~\cite{Warnick:2013hba}, we expect this approach to be valid in generic asymptotically AdS spacetimes.

We have investigated the behaviour of $\tau$ for Gaussian and tanh quenches with the idea of associating this quantity to a well defined notion of initial time. Our results are positive, although some unsolved questions remain. We find that the qualitative dependence of $\tau$ on the abruptness of the quench and $\kappa B$ matches the expectations: $\tau$ becomes smaller close to the adiabatic regime or for higher $\kappa B$.

However, it is still to be investigated how $\tau$ depends on the precise quench protocol and whether $\tau$ can be used to identify a generic classification for the quenches. In particular, the Gaussian quench showed a rather unexpected behaviour at $\kappa B = 0$, namely the time scale $\tau$ appeared to be independent of the width. This issues deserve deeper analysis and we leave it to future work.    

In addition, we have studied the behaviour of the dual current operator for fast quenches (motivated by \cite{Buchel:2013lla}) where a universal response regime was found in a holographic model with a dual scalar operator. Our results show that for fast quenches the quantity $\delta$ defined in equation \eqref{eq:delta} shows a universal behaviour: it is independent of the magnetic field and it scales with a fixed exponent with the width of the quench. Moreover, we have shown that this universal regime is suppressed the higher $\kappa B$, i.e. one needs to perform faster quenches to reach the universal regime the higher $\kappa B$.

Our results for both $\tau$ and $\delta$ indicate that the relaxation time of the system is smaller with increasing $\kappa B$. This is coherent with the notion of $\kappa B$ being a coupling of the current to the electric field\footnote{This becomes apparent if one considers the negative magnetoresistance in the case of parallel external fields $J\sim (\kappa B)^2 E$.}.  

The last main point of this work is the emphasis on the dynamics of the current for late times. By explicitly computing the fate of the current after quenches in the electric field, we have shown that oscillatory, long-lived currents are produced in presence of the anomaly and strong enough magnetic fields in our holographic model. By computing the QNMs of the system with backreaction we have been able to show that this is a consequence of the presence of Landau levels\footnote{It is worth mentioning that Landau level resonances have been experimentally found in Dirac semimetals \cite{2016arXiv160102316Y, PhysRevLett.115.176404}.}. This has been done by looking at the dependence of the lowest QNMs on $\kappa$ and $B$. We have observed that for non-zero $\kappa$ the real part of the frequency goes as  $\sim\sqrt B$ for large $B$, pointing towards the appearance of Landau levels for $\kappa>0$. As already discussed in the main text, the fact that the characteristic $\sqrt{B}$ behaviour is absent for $\kappa =0$ indicates that no dual fermionic d.o.f. are charged under the symmetries in consideration when the Chern-Simons term is not present in the bulk, i.e. when there is no dual anomaly.

 Whether resonances are to be found in the conductivity depends not only on the magnetic field, but on $\kappa$ as well. As we saw, the imaginary part of the quasi normal frequencies asymptotes the real axis only if $\kappa \gtrsim 1/2$. It is tempting to speculate whether this is related to the existence of a quantum critical point for $\kappa>1/2$ studied in \cite{D'Hoker:2010rz, D'Hoker:2010ij, Ammon:2016szz}. 
 
In addition, we explored the idea of directly exciting the Landau resonances by means of an oscillatory external electric field finding a resonant pattern as expected\footnote{This response was already implicit in \cite{Jimenez-Alba:2014iia, Sun:2016gpy}.}. Moreover it would be interesting to check whether long lived oscillating currents can be generated by means of an electric quench in Weyl semimetals. A more realistic treatment of this would require to introduce some (``intrinsic'') mechanism for axial charge relaxation \cite{Jimenez-Alba:2014iia}. We propose to check this statements in the weak coupling regime. Regarding this resonances, it is worth mentioning that the system shows several similarities to holographic p-wave superconductors. Both present dissipationless transport driven by a broken symmetry and both cases \cite{Ammon:2010pg} present resonances at finite frequencies in the anisotropic direction. It can be interesting to explore this analogy in more depth. 

Let us finally mention that initially we had Weyl/Dirac semimetals in the back of our heads, which is described at strong coupling by \cite{Landsteiner:2015lsa}. However, we have considered a simpler model which we think captures the core properties we wanted to focus on. It would be interesting to make an analogous study with the holographic model~\cite{Landsteiner:2015lsa}. 

\section*{Acknowledgements}
 \noindent We thank Marcus Ansorg, Birger Boening, Markus Gardemann, Holger Gies, Markus Heinrich, Carlos Hoyos, Matthias Kaminski, Karl Landsteiner, Julian Leiber, Chris Rosen, Andrej Rostworowski, Jackson Wu and Ho-Ung Yee for fruitful discussions. A.J.~and R.M.~acknowledge financial support by \textit{Deutsche Forschungsgemeinschaft (DFG)} GRK 1523/2. R.M.~was supported by CNPq under the programme "Ci\^encia sem Fronteiras". The work of L.M. is supported by the ERC Advanced grant No.339140 Gravity, Black Holes and Strongly Coupled Quantum Matter.
\begin{appendix}

\section{Pseudospectral methods}\label{sec:methods}\noindent
 In this section, we give more details on the numerical techniques used in this work. As already mentioned along the text, all the relevant equations are solved with spectral methods. These methods are widely used for boundary-value problems~\cite{Boyd00,canuto_2006_smf} (typically elliptic equations) and they find several applications for time-independent problems within numerical relativity~\cite{grandclement_2009_smn} and numerical holography~\cite{Dias:2015nua}. Recently, there has been also some development in the application of spectral methods along the time direction in dynamical scenarios~\cite{Macedo:2014bfa}.
 
Here, we discuss the main features of the method with emphasis on its application for eigenvalue problems and for systems of differential equations with extra parameters. Then, we review the algorithm for the fully spectral code introduced in~\cite{Macedo:2014bfa}.

 \subsection{The eigenvalue problem: quasi-normal modes}\label{sec:SpecMeth_QNM}\noindent
 For convenience, let us reproduce here the general representation of the ordinary differential equation characterising the QNM as we introduced in \eqref{eq:QNMEq} (for a fixed background) or \eqref{eqS::gevp} (for the backreacted system)
 \[
 {\boldsymbol \alpha} [\phi_n] + s_n {\boldsymbol \beta} [\phi_n]  = 0.
\]
This equation is to be solved in the domain $\rho\in[0,1]$. Due to the choice of the ingoing Eddington-Finkelstein coordinates to the background metric, we obtain only a linear term in the complex parameter $s_n$ characterising the QNM\footnote{Recall that $s$ is the Laplace parameter, which is related to the Fourier frequency $\omega$ by $s=-i\omega$.}. Furthermore, thanks to this coordinate choice, the surfaces of constant time penetrate the black-hole horizon and therefore the ingoing boundary condition on the horizon is automatically realised here by the geometry, i.e., they correspond to the regularity condition of  \eqref{eq:QNMEq} at $\rho=1$. In the same way, equation~\eqref{eq:QNMEq} provides us with the proper regularity conditions at $\rho=0$, so no further information is needed for the solution of equation~\eqref{eq:QNMEq}.    

In order to discretise equation~\eqref{eq:QNMEq}, we fix a numerical resolution $N_\rho$ and we introduce a Chebyshev-Lobatto grid in the $\rho$ direction via
\beq
\label{eq:LobattoGrid}
\rho_j = \frac{1}{2}\left[ 1+\cos\left(\pi \frac{j}{N_\rho} \right) \right] \quad j = 0, \ldots, N_{\rho}.
\eeq
Then, we recall that ${\boldsymbol \alpha(\rho)}$ and ${\boldsymbol \beta(\rho)}$ are (second/first order, respectively) differential operators acting on the radial coordinate $\rho$. Hence, a discrete representation $\hat{\alpha}$ and $\hat{\beta}$ is obtained by substituting the differential operators $\partial_\rho$ and $\partial^2_{\rho \rho}$ by the discrete Chebyshev-Lobatto spectral differentiation matrices $\hat{D}_{\rho}$ and $\hat{D}^2_{\rho \rho} = \hat{D}_{\rho}\cdot \hat{D}_{\rho}$ , whose expression can be found in~\cite{canuto_2006_smf}. In its discrete form, equation~\eqref{eq:QNMEq} has the structure of a generalised eigenvalue problem and both eigenvalues $s_n$ and eigenvectors $\vec{\phi}_n$ are easily obtained with \texttt{Mathematica}. 

First thing to observe is that we obtain a total of $N_\rho+1$ eigenvalues and eigenvectors. Yet, not all of them are trustful numerical solution. It is crucial that one performs a convergence test to determine which of the obtained solutions are stable and converge to fixed value as one increases the resolution $N_\rho$. Our empirical observation shows that for a given $N_{\rho}$, the first $n_\text{QNM} \sim \sqrt{N_\rho}$ QNMs correspond to a reliable numerical solution. In order to keep high-accurate solutions, we fix $N_{\rho} = 300$ and considered only the first $n_\text{QNM} = 12$ quasi-normal modes in this work. Secondly, we would like to stress that we are also interested in the eigenfunctions $\vec{\phi}_n$, since they are needed in the calculation of the QNM amplitudes $\eta_n$. Each component of $\vec{\phi}_n$ corresponds to the value of the function $\phi(\rho)$ at the grid point, i.e.,~$\phi_n^j \equiv  \phi(\rho_j)$. Notice that the eigenfunctions are uniquely defined up to a normalisation constant. Thus, we work with the conveniently rescale quantity\footnote{From now on, we omit the notation $\bullet^{\rm norm}$ and $\phi(\rho)$ is always the normalised quantity.} $\vec{\phi}_n^{\rm norm} = \vec{\phi}_n/\phi_{n}^{N_{\rho}}$ which give us $\phi^{\rm norm}_n(0) = 1$.
 
  \subsection{Equation with a free parameter: QNM amplitudes}\label{sec:SpecMeth_QNM_Amplitudes}\noindent
  We now proceed with the discussion of the solution of equation~\eqref{eq:Amplitude} (which we reproduce here once more for convenience)
\[
{\mathbf A} (s_n) [ \bar{W} ] + \eta_n \, {\boldsymbol \beta}[\phi_n] = \bar{Q}.
\]
We remind that ${\mathbf A} (s_n) = {\boldsymbol \alpha} [\phi_n] + s_n {\boldsymbol \beta} [\phi_n]$. Moreover, we have simplified the source term on the r.h.s.~with the introduction of $\bar{Q}={\boldsymbol \beta} [U_{\rm in}] - \bar{S} $. 

Our objective is to numerically solve equation~\eqref{eq:Amplitude}  for {\em both} $W(\rho)$ and $\eta_n$. We also make use of spectral methods for this task so, in principle, the discretisation procedure in terms of the Chebyshev-Lobatto grid points~\eqref{eq:LobattoGrid} and the discrete Chebyshev-Lobatto spectral differentiation matrices is the same as described in \ref{sec:SpecMeth_QNM}.

Note, however, that we have now a total of $N_\rho +2$ unknown variables, which we collect into a single vector (with $W^i \equiv W(\rho_i)$)
\beq
\vec{X}^{\rm T} = \left( W^0, \,\ldots\,, W^{N_\rho}, \, \eta_n\right).
\eeq
The discrete version of equation~\eqref{eq:Amplitude} provides us only with $N_\rho +1$ equations, though. The extra condition needed to complete the system is obtained after observing that the solution $\bar{W}(\rho)$ is not unique. Assuming $\bar{W}_{\rm a}(\rho)$ is solution to equation~\eqref{eq:Amplitude}, then, $\bar{W}_{\rm b}(\rho) = \bar{W}_{\rm a}(\rho) + C \phi_n(\rho)$ also satisfies \eqref{eq:Amplitude}. Therefore, we must impose an extra normalisation condition to $W(\rho)$, which here we fix as 
\beq
\label{eq:NormW}
W(0) = 1.
\eeq 
Taking into account the discretised spectral representation of equation~\eqref{eq:Amplitude} together with the condition \eqref{eq:NormW}, and introducing the notation $\vec{\beta}_\phi = \hat\beta \cdot \vec{\phi}$, we end up with the linear system $\hat{M}\cdot \vec{X} = \vec{q}$, whose components can be explicitly expressed as
\beq
\label{eq:QNMAmplit_LinSyst}
\left(
\begin{array}{cc}
A^{i}{}_{j} & \beta_{\phi}^i \\
\delta^{j}{}_{N_\rho} & 0
\end{array}
\right)
\left(
\begin{array}{c}
W^{j} \\
\eta_n
\end{array}
\right)=
\left(
\begin{array}{c}
Q^{i} \\
1
\end{array}
\right), \quad i = 0,\ldots, N_\rho.
\eeq 
We observe that $\eta_n$ does not depend on the normalisation $W(\rho_0) = W_0$.
 
While the algorithm used to solve equation~\eqref{eq:Amplitude} is essentially the one described above,  we must face a caveat introduced by the logarithmic terms presented in the source term $\bar{Q}(\rho)$ [see equations~\eqref{eq:Source_ak} and \eqref{eq:LapTransSource}], which leads to an algebraic convergence rate of the spectral scheme. In fact, for generically non-vanishing boundary data $V_0(v)$, the source $\bar{Q}(\rho)$ is merely ${\cal C}^0$ due to the term $\sim \rho\log(\rho)$ and in practice, one faces difficulty in finding the amplitudes $\eta_n$ already for $n\gtrsim 2$.

To overcome this problem, we introduce a new coordinate~\cite{Kalisch:2016fkm} $z\in[0,1]$ via
\beq
\rho = \e^{1-{1}/{z}},
\eeq
which allows one to map the problematic ${\cal C}^{k-1}$ terms $\sim\rho^{k} \log(\rho)$ into the ${\cal C}^\infty$ expressions $\sim(1-1/z) e^{(k-k/z)}$. By rewriting\footnote{The point $z=0$ must be treated with care. In this limit, one obtains ${\cal W}_{,z}(0) = 0$ and this property should be explicitly implemented when constructing the matrix $\hat{M}$.} equation~\eqref{eq:Amplitude} in terms of $z$ and discretising the system with Chebyshev-Lobatto grid points $z_i$ together with differentiation matrices $\hat{D}_z$, we obtain a spectral convergence rate\footnote{Here, spectral convergence means that the convergence rate is faster than algebraic. However, the rate does not decay exponentially since ${\cal W}(z)$ is not analytic at $z=0$.} for the solution ${\cal W}(z) = W(\rho(z))$ and therefore a much more stable scheme for finding the amplitudes $\eta_n$.

A second remark regarding the efficiency of the code is related to the presence of rather huge numbers ($\sim 10^{30}-10^{50}$). Such values are a consequence of the Laplace-transformation of the boundary function $\bar{V}_0(s_n)$, which contributes significantly to the source function $\bar{Q}(\rho)$ (see \eqref{eq:LapTransSource}). Therefore, it is also convenient to rescale the vectors in \eqref{eq:QNMAmplit_LinSyst} by $\vec{q} = \bar{V}_0(s_n) \vec{p}$ and $\vec{X} = \bar{V}_0(s_n) \vec{Y}$ and solve the system $\hat{M}\cdot \vec{Y} = \vec{p}$.

In order to obtain the desired high accuracy for the first $n_{\rm QNM} = 12$ quasi-normal modes amplitudes $\eta_n$, we set the resolution to $N_z = 600$. It is true that both resolutions $N_\rho=300$ (for the solution of the eigenvalue problem - sec.~\ref{sec:SpecMeth_QNM}) and $N_z=600$ are quite extreme for codes based on spectral methods. We mention however, that for a given physical parameter $\kappa B$ the generalised eigenvalue problem and the inversion of the matrix $\hat{M}$ must be performed only once. The results can be saved and applied afterwards several times for any r.h.s $\vec{q}$. Still, a more sophisticated and efficient approach based on the work~\cite{Ansorg:2016ztf} together with a systematic study of the QNM amplitudes in a more generic context is planned to be presented in a forthcoming article. 
 \subsection{Hyperbolic equation: fully spectral code}\label{sec:TimeSpec}\noindent
We end this section with a brief discussion of the fully spectral code used to solve the equation \eqref{eq:DynEq_TimeDom} in terms of the coordinates $\{v,\rho\}$.  A detailed description of the method can be found in~\cite{Macedo:2014bfa}. Let us start by assuming we are looking for a solution in a generic time interval $v\in [v_{\rm a}, v_{\rm b}]$. Given the initial data $U_{\rm a}(\rho)$ at $v=v_{\rm a}$, we introduce the auxiliary fields $P(v,\rho)$ via
\beq
\label{eq:AuxFields}
U(v,s) = U_{\rm a}(\rho) + (v - v_{\rm a})P(v, \rho).
\eeq
For prescribed numerical resolution $N_\rho$ and $N_v$, we work with the Chebyshev-Lobatto grid in the $\rho$-direction given in equation~\eqref{eq:LobattoGrid} and with the Chebyshev-Radau collocation points in the $v$-direction
\beq
\label{eq:grid}
v_k = \frac{ v_{\rm b} + v_{\rm a}}{2} +  \frac{v_{\rm b} - v_{\rm a}}{2} \, \cos\left(\frac{2k\pi}{2N_v+1}\right),\,\, 
k=0...N_v,
\eeq
The function values $V_{ki}=V$ are stored in a vector $
\vec{X}^{\rm T} = \left( V_{ki} \right)_{k=0\dots N_{v}, \, i = 0 \dots N_{\rho}},$ from which Chebyshev coefficients $c_{kj}$ of the field $P(v, \rho)$ are computed by inverting the equations  
\bea
\label{eq:SpecAppr}
P(v_k,\rho_i) &=& \sum_{n=0}^{N_u}  \sum_{l=0}^{N_\rho}  
c_{nl} T_n\left(\frac{2v_k-v_{\rm a}-v_{\rm b}}{v_{\rm b}-v_{\rm a}}\right)
          T_l\left(1-2\rho_i\right),
\eea
with $T_j(\mu) = \cos[j\arccos(\mu)]$ the Chebyshev polynomials of the first kind. After calculating spectral approximations of the fields' derivatives, equation~\eqref{eq:DynEq_TimeDom} yields a linear algebraic system, which is solved with the iterative BiCGStab method. Furthermore, we also provide the BiCGstab method with a pre-conditioner based on a Singly Implicitly Diagonally Runge-Kutta (SDIRK) method~\cite{Macedo:2014bfa}.

After obtaining the solution $U(v, \rho)$ for $v \in [v_{\rm a}, v_{\rm b}]$, the values $U(v_{\rm b}, \rho)$ at the upper time boundary $v_{\rm b}$ serve as initial data for a subsequent time domain $v\in [v_{\rm b}, v_{\rm c}]$. This procedure allows to divide the whole time interval $v\in[0,v_{\rm final}]$ into smaller sub-intervals, with the very first one being $[0, v_{\rm a}]$. The size of each time interval $\Delta v$ can be chosen according to the quench profile. If $\Delta$ is a characteristic time length for a given quench, then we fix $\Delta v = \Delta/4$. Furthermore, we set the resolution in the time direction to $N_v \sim 25$. The radial direction requires a higher resolution $(N_\rho \sim 100)$ due to the presence of logarithmic terms in the source function $S(v,\rho)$.

 \section{Schr\"odinger type analysis for the QNMs}\label{sec:QNMapp}\noindent
To understand the behaviour of the QNMs approaching the real axis in the complex $\omega$ plane in the probe limit, we rewrite the perturbation equation in the form 
\begin{align}
\label{eqp} (\partial^2 +\omega^2 -V_{\rm eff}) \phi =0\,.
\end{align}
In order to do so we first change coordinates
\begin{align}
\dd s^2= -U(r) \dd t^2 +\frac{\dd r^2}{U(r)} + r^2\,(\dd x^2+\dd y^2+\dd z^2)\,.
\end{align}
 where $U(r)= r^2-1/r^2$ is the blackening factor and we restrict ourselves to the probe limit. In this coordinates the boundary is located at $r\rightarrow\infty$. We now consider the e.o.m. to first order in perturbations $\delta V_j(r,t,x)=v_j(r) \e^{-\im \omega t+ \im k x}$, $\delta A_j(r,t,x)=a_j(r) \e^{-\im \omega t+ \im k x}$ on top of the magnetic field background in the $k=0$ limit. Since the system is linear, this just gives rise to the background equations (\ref{eq:eom1},\ref{eq:eom2},\ref{eq:eom3}). Using the constraint equation one finds 
\begin{align}\label{eq:luis1}
v_z''+v_z'\left(\frac{U'(r)}{U(r)}+\frac{1}{r}\right)+\frac{r^4 \omega ^2-144\, (\kappa B )^2\, U(r)}{r^4\, U(r)^2}\, v_z=0\,,
\end{align}
where the prime denotes radial derivative and $v_z =v_z(r)$.
To attain the form \eqref{eqp}, we define
\begin{align}
\frac{\dd\tilde{r}}{\dd r} \equiv H(r)\,, \hspace{4cm}
v_z(r) =  S(r) \tilde{v}_z(r)\,,
\end{align}
using the dot $\dot{\ }$ to denote $\partial_{\tilde{r}}$ \eqref{eq:luis1} can be rewritten as
\begin{align}
&\ddot{\tilde{v}}_z +\left(\frac{2 S'}{S H} + \frac{1}{H}\left(\frac{U'}{U}+\frac{1}{r}\right)  + \frac{ H'}{H^2}\right)\, \dot{\tilde{v}}_z \nn \\ &\quad + \left(\frac{S'}{SH^2}\left(\frac{U'}{U}+\frac{1}{r}\right) +\frac{S''}{SH^2}  -\frac{144 \, (\kappa B )^2\,}{r^4U H^2} +\frac{\omega^2}{H^2U^2} \right)\, \tilde{v}_z=0\,.\label{finaleq}
\end{align}
Imposing this expression to match with \eqref{eqp} fixes $S$ and $H$ up to an unimportant constant and leads to 
\begin{align}
 \left(\partial_{\tilde r}^2+\omega ^2-\frac{U(r) \left(288 \, (\kappa B )^2\,+r^3 U'(r)\right)}{2 r^4}+\frac{U(r)^2}{4 r^2}\right) \tilde{v}_z=0\,.
\end{align}
The explicit expression effective potential is
\begin{align}\label{eq:effpot}
V_{\rm eff} (\kappa^2 B^2;r) = \frac{\left(r^4-1\right) \left(576 \,(\kappa B)^2+3 r^4+5\right)}{4 r^6}\,.
\end{align}
Note that $\lim_{r\rightarrow \infty} V_{\rm eff} = \infty$, whereas $V_{\rm eff} (r=1)=0$. 
 \begin{figure}[t!] 
\centering
\includegraphics[width=6.9cm]{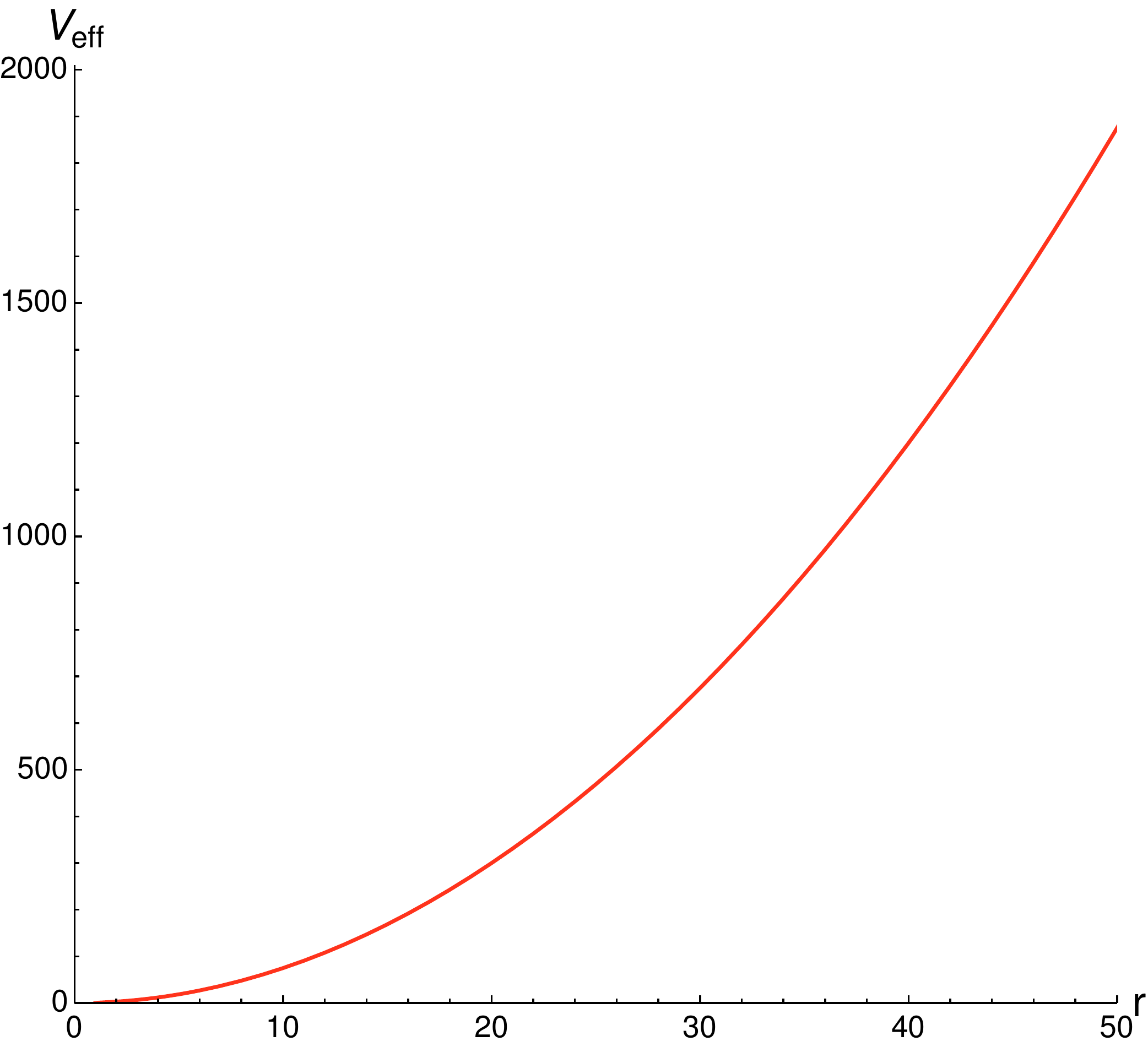}
\hspace{1cm}
\includegraphics[width=6.9cm]{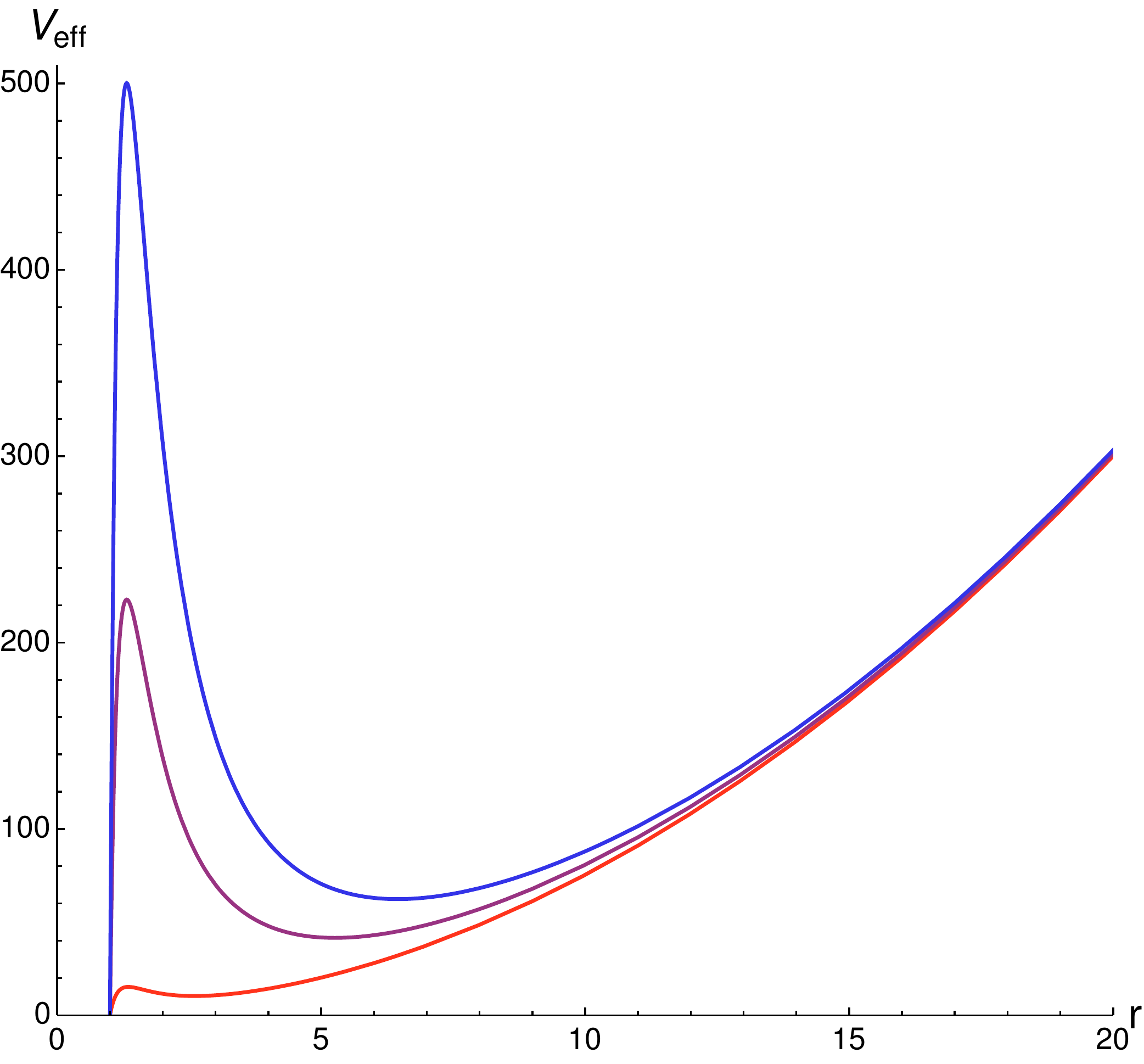}
\caption{\label{fig:potential} Left: Effective potential (see equation \eqref{eq:effpot}) for $\kappa B=0$ against $r$. The curve grows monotonically with increasing $r$. Right: Effective potential for $\kappa B=0.1,1,2$ (red-blue) against $r$. The $\kappa B$ term becomes relevant close to the horizon and forms a higher barrier for increasing $\kappa B$. }
\end{figure}
We plot this function in Figure \ref{fig:potential}. As explained in \cite{Kiritsis:2015oxa} the $\kappa B=0$ case is compatible with a conductivity displaying a continuous spectrum due to the fact that the potential is unbounded. However, as $\kappa B$ increases the effective potential develops a barrier close to the horizon, creating more bound states as $\kappa B$ grows bigger. For $\kappa B \rightarrow \infty$ the barrier is infinite and we have a binding potential: there is only a discrete set of allowed frequencies and the QNMs do not display an imaginary part because they cannot reach the black hole due to the infinite barrier. 

\section{Quasi Normal Modes}\label{app:QNM}\noindent
\noindent
As explained in the main text, in order to be able to vary $\kappa$ and $B$ independently we have to take backreaction into account (see e.g. \cite{D'Hoker:2009mm}). Since the magnetic field breaks rotational invariance we choose the following ansatz for the metric
\begin{equation}
 \dd s^2=\frac{1}{\rho^2}\left(-U(\rho)\, \dd v^2 - 2\,\dd v\,\dd \rho+W(\rho)^2\,(\dd x^2+\dd y^2)+H(\rho)^2\,\dd z^2\right).
\end{equation}
The equations of motion for the background in the so called trace reduced form read
\begin{equation}
 R_{\mu\nu}=-4\,g_{\mu\nu}+\tau^2\,\left(-\frac 16\, g_{\mu\nu}\,\left( F_{\alpha\beta}F^{\alpha\beta}+H_{\alpha\beta}H^{\alpha\beta}\right)+g^{\alpha\beta}\,\left(F_{\mu\alpha}F_{\nu\beta}+H_{\mu\alpha}H_{\nu\beta}\right)\right).\label{eqs:efg}
\end{equation}
Due to diffeomorphism invariance we can set the black hole horizon to $\rho=1$. The metric function $U(\rho)$ is the blackening factor and hence has to vanish at the horizon $U(1)=0$.
Before solving the equations of motion we first consider the asymptotic expansions. Imposing asymptotically AdS
\begin{equation}
 U'(0)=0, \ W(0)=1, \ H(0)=1
\end{equation}
leads to
\bea
 U(\rho) & =1+\rho^4 \,\left[\mathbf{u}_4+\mathcal O(\rho^2)\right]+\rho^4\,\log(\rho)\,\left[\frac{B^2 \tau^2}{3}+\mathcal O(\rho^2)\right],\label{uexp}\\
  W(\rho) & =1+\rho^4 \,\left[-\frac{\mathbf{h}_4}{2}+\mathcal O(\rho^2)\right]+\rho^4\,\log(\rho)\,\left[-\frac{B^2 \tau^2}{12}+\mathcal O(\rho^2)\right],\label{wexp}\\
   H(\rho) & =1+\rho^4 \,\left[\mathbf{h}_4+\mathcal O(\rho^2)\right]+\rho^4\,\log(\rho)\,\left[-\frac{B^2 \tau^2}{6}+\mathcal O(\rho^2)\right],\label{hexp}
\eea
where we set the term linear in $\rho$ in equation \eqref{uexp} (and therefore in equations \eqref{wexp} and \eqref{hexp}) to zero in order to fix the remaining diffeomorphisms. Furthermore we can fix $2\tau^2=1$ since it appears in the equations always as a product with the magnetic field $B$.

Using the ansatz $V_m=V_y(x)=Bx$ and $A_m=0$ we can solve the equations of motion by a spectral method. To do so we first rescale our functions properly using the expansions equation \eqref{uexp}-\eqref{hexp} and solve for given $\kappa$ and $B$ equation \eqref{eqs:efg} for the rescaled functions.

With the background solutions at hand we can look at the fluctuations (c.f. \cite{Janiszewski:2015ura}). Since the factor proportional to the Chern-Simons coupling is metric independent the metric fluctuations will decouple from the fluctuations of the gauge fields. Furthermore the (1) and (2) sector decouple as well and we have just a coupling of the (0)-(3) sector. Setting the momentum
$k$ to zero, we notice, that only the $a_0$-$v_3$ and $a_3$-$v_0$ components are coupled.
For the former the constraint equation reads
\begin{equation}
\dot a_0'(v,\rho)+\frac{12\, \kappa B \,\rho}{H(\rho)\, W(\rho)^2}\, \dot v_3(v,\rho)=0.
\end{equation}
Similar to the non-backreacted case we can integrate the equation in time and obtain
\begin{equation}
 a_0'(v,\rho)=-\frac{12\, \kappa B\, \rho}{H(\rho)\, W(\rho)^2}\, v_3(v,\rho)+C_1.
\end{equation}
Like before we set the constant $C_1$ to zero. With this relation between $a_0'$ and $v_3$ we can eliminate $a_0$ in the $v_3$ equation and obtain after a Laplace transformation 
\begin{equation}
 {\boldsymbol \alpha}[v_{3,n}]+s_n\,{\boldsymbol \beta}[v_{3,n}]=0, \label{eqS::gevp}
\end{equation}
with 
\begin{align}
  \boldsymbol{\alpha}&=-(12\,\kappa B )^2\, \rho^3\,H(\rho)+\rho\, H(\rho) U(\rho) W(\rho)^4 \,\frac{\dd^2}{\dd \rho^2}\nonumber\\&+\left[H(\rho) W(\rho)^3 \left(2 \rho \,U(\rho) W'(\rho)\!-\!W(\rho) (U(\rho)-\rho\,U'(\rho)\right)-\rho\, U(\rho) W(\rho)^4 H'(\rho)\right]\,\frac{\dd}{\dd \rho}
\intertext{and}
\boldsymbol{\beta}&= \left[H(\rho) W(\rho)^3 \left(W(\rho)-2 \rho\, W'(\rho)\right)\right] +2 H(\rho) W(\rho)^4 \,\rho\,\frac{\dd}{\dd \rho}\,.\end{align}
We solve this equation by means of pseudospectral methods imposing the non-normalizable mode to zero. The QNMs we are interested in converge slowly for high values of the magnetic field. To ensure the convergence we improve the spectral solution by a coordinate mapping of the radial variable $\rho\mapsto\tilde\rho^2$ which moves the gridpoints in the direction of the boundary localised at $\tilde\rho=0$. Furthermore we use a grid with $N_\rho=200$.
\end{appendix}

\newpage
\bibliographystyle{JHEP}
\bibliography{paper}

\end{document}